\documentclass[12pt]{article}

\usepackage{mathrsfs}
\usepackage{fullpage}
\usepackage{amsfonts}
\usepackage{graphicx}
\usepackage{mathrsfs}
\usepackage{amsmath}
\usepackage{amssymb}
\usepackage{float}
\usepackage{subfig}
\usepackage{rotating,color}
\usepackage{xcolor,natbib,epstopdf}
\usepackage{appendix,algorithm,algpseudocode}
\usepackage{hyperref}
\hypersetup{
	colorlinks=true,    
	linkcolor=blue,    
	citecolor=blue,    
	urlcolor=blue      
}
	\renewcommand\appendix{\par
	\setcounter{section}{0}
	\setcounter{subsection}{0}
	\setcounter{lemma}{0}
	\setcounter{equation}{0}
	\renewcommand{\thelemma}{\arabic{lemma}}
	\gdef\thesection{Appendix \Alph{section}}
	\gdef\thesubsection{\Alph{section}.\arabic{subsection}}}
\renewcommand\arraystretch{0.8}

\newcommand{\be}{\begin{equation}}
\newcommand{\ee}{\end{equation}}
\newcommand{\beaa}{\begin{eqnarray*}}
\newcommand{\eeaa}{\end{eqnarray*}}
\newcommand{\bea}{\begin{eqnarray}}
\newcommand{\eea}{\end{eqnarray}}

\newcommand{\eq}[1]{$(\ref{#1})$}

\newtheorem{theorem}{ \noindent T{\footnotesize HEOREM}}
\newtheorem{prop}{ \noindent P{\footnotesize ROPOSITION}}[section]

\newtheorem{remark}{ \noindent R{\footnotesize EMARK}}[section]

\newtheorem{example}{ \noindent E{\footnotesize XAMPLE} }[section]

\def\diag{\mathrm {diag}}

\newcommand{\bm}{\boldsymbol}

\def\R{{\bf R}}
\def\I{{\bf I}}

\def\P{{\bf M}}

\def\Y{{\bm Y}}
\def\f{{\bm f}}

\def\D{{\bf D}}

\def\bms{{\bm\Sigma}}

\def\cp{\mathop{\rightarrow}\limits^{p}}
\def\cd{\mathop{\rightarrow}\limits^{d}}
\def\mR{\mathbb{R}}

\def\bmv{\bm \varepsilon}

\def\boxit#1{\vbox{\hrule\hbox{\vrule\kern6pt  \vbox{\kern6pt#1\kern6pt}\kern6pt\vrule}\hrule}}
\def\bse{\begin{eqnarray*}}
\def\ese{\end{eqnarray*}}
\def\be{\begin{eqnarray}}
\def\ee{\end{eqnarray}}
\def\bsq{\begin{equation*}}
\def\esq{\end{equation*}}
\def\bq{\begin{equation}}
\def\eq{\end{equation}}

\def\mR{\mathbb{R}}
\def\n{\nonumber}

\def\diag{\mbox{diag}}

\def\D{{\bf D}}

\def\I{{\bf I}}

\def\R{{\bf R}}

\def\Y{{\bf Y}}

\def\diag{\hbox{diag}}

\def\diag{\hbox{diag}}

\def\squarebox#1{\hbox to #1{\hfill\vbox to #1{\vfill}}}

\def\0{{\bf 0}}
\def\1{{\bf 1}}

\def\mR{\mathbb R}

\def\diag{\hbox{diag}}

\def\diag{\hbox{diag}}

\allowdisplaybreaks

\title
{\bf Robust Spatial-Sign-Based Testing of High-Dimensional Alpha in Conditional Factor Models}
\author{Ping Zhao$^{1}$, Hongfei Wang$^{2}$\\
	$^{1}$School of Mathematical Sciences, Tiangong University\\
	$^{2}$Department of Statistics, Nanjing Audit University}
\date{}

\begin{document}
\maketitle

\begin{abstract}
	This paper develops a new framework for alpha testing in high-dimensional factor pricing models with time-varying coefficients. To detect sparse alternatives, we propose a spatial-sign-based max-type test and derive its limiting null distribution. A key theoretical result is that our statistic is asymptotically independent of the spatial-sign-based sum-type test proposed by \cite{zhao2023robust}. Exploiting this independence, we construct an adaptive testing procedure via the Cauchy combination method. This approach integrates the complementary strengths of both max-type and sum-type statistics, ensuring robust power across diverse sparsity levels. Extensive simulations and an empirical application demonstrate that the proposed test is resilient to heavy-tailed distributions and maintains superior performance under various alternative specifications.
\end{abstract}


\section{Introduction}

Testing the null hypothesis that asset-pricing alphas are jointly equal to zero is a central task in empirical finance \citep{gibbons1989test,fama2015five}. Such tests provide a formal basis for evaluating the empirical adequacy of factor pricing models and inform both model selection and portfolio construction. In modern empirical settings, researchers increasingly encounter two significant challenges. First, datasets are frequently high-dimensional \citet{fan2015power,gu2021autoencoder}. The number of assets $N$ in the cross-section may exceed or be comparable to the time-series dimension $T$. In such cases, the sample covariance matrix can become singular, rendering the classical GRS test infeasible or unreliable. Second, asset returns often exhibit heavy-tailed distributions and other non-Gaussian features \citet{mandelbrot1963variation,fama1965behavior,cont2001empirical}, which may induce size distortions and power loss in testing procedures derived under sub-Gaussian or light-tailed assumptions. These features motivate the development of testing procedures that remain valid in high-dimensional settings and are robust to heavy-tailed distributions. 

A further complication arises from potential time variation in factor exposures. While unconditional factor models typically assume constant factor loadings, this assumption may be violated over long sample periods due to structural shifts in the economic environment and evolving market conditions. Empirical evidence, including \cite{ghysels1998stable} and \cite{lewellen2006conditional}, suggests that both alphas and factor loadings can vary over time. Ignoring such dynamics may lead to model misspecification and biased inference on alpha, potentially resulting in incorrect acceptance or rejection of a factor model. Consequently, global tests of alpha within conditional factor models, which explicitly accommodate time-varying parameters, have attracted increasing attention \citep{ferson1999conditioning, fu2025distinguishing}.

As \citet{li2011testing} observed, testing whether every asset's alpha is zero at every point in time is an exceptionally stringent requirement that is frequently rejected in practice. Accordingly, much of the literature focuses on the global null hypothesis that the time-average of the conditional alpha is zero. \cite{li2011testing} and \cite{ang2012testing} proposed nonparametric Wald-type tests for this purpose, modeling both alphas and factor loadings as smooth functions of time. However, the validity of these tests typically relies on an asymptotic framework in which $N$ is fixed as $T \to \infty$. In high-dimensional settings where $N$ is large relative to $T$, the estimated covariance matrix often becomes singular or poorly conditioned, causing these conventional Wald-type tests to fail.

To address the challenges posed by high-dimensional data, a growing body of research has developed testing procedures for conditional factor models that allow $N$ to diverge with $T$. These methods generally fall into two categories based on their power properties: sum-type tests, which are typically more powerful against dense alternatives where many alphas are non-zero, and max-type tests, which are designed for sparse alternatives where only a few alphas are non-zero. For example, \cite{ma2020testing} developed a sum-type statistic that accommodates $N \gg T$ under time-varying settings, but its power is inherently limited when the deviation from the null is sparse. Conversely, \cite{zhao2023robust} proposed a robust sum-type spatial-sign test, and \cite{zhang2025maximum} introduced an max-type test based on the maximum of squared test statistics. A common limitation of these approaches is that each is optimized for a specific regime, either dense or sparse, and may suffer from substantial power loss when the true nature of the alternative hypothesis is unknown.

Recently, \cite{ma2024adaptive} proposed an adaptive testing procedure by combining max-type and sum-type statistics, leveraging their asymptotic independence under the null hypothesis. While this approach is more flexible, its theoretical validity relies on sub-Gaussian tail assumptions. Under heavy-tailed distributions, the moment conditions required for the asymptotic distributions of these statistics may fail to hold, leading to a potential loss of power or size distortions. Therefore, constructing a nonparametric testing procedure for conditional factor models that is both adaptive to the sparsity of the alternative and robust to heavy-tailed distributions remains an important and unresolved problem in the literature.

To address these challenges, this paper develops a robust and adaptive testing framework for alpha in conditional time-varying factor models. We construct a new max-type statistic tailored to the conditional time-varying setting and establish its asymptotic behavior. Crucially, we prove the asymptotic independence between this max-type statistic and a complementary sum-type statistic under the null hypothesis. Leveraging this asymptotic independence, we combine the two statistics into a single adaptive test that overcomes the limitations of existing adaptive methods such as \cite{ma2024adaptive}, whose theoretical guarantees rely on sub-Gaussian tail assumptions. We evaluate the finite-sample performance of the proposed test through Monte Carlo experiments and an empirical application.

The main contributions of this paper are threefold:

First, we relax the stringent light-tailed or sub-Gaussian requirements prevalent in existing high-dimensional alpha tests. By establishing asymptotic theory under substantially milder moment and dependence conditions, the proposed procedure ensures robustness in heavy-tailed environments, a feature frequently observed in empirical asset returns but often overlooked by conventional tests.

Second, we develop a new max-type statistic specifically tailored for conditional time-varying factor models. Compared with the sum-type test in \cite{zhao2023robust}, the proposed max-type test is more suitable for sparse alternatives and exhibits superior power relative to existing max-type methods under such regimes.

Third, we demonstrate the asymptotic independence between the sum-type and max-type statistics under the null hypothesis. Leveraging this result, we construct an adaptive combination test that attains superior power against both dense and sparse alternatives while simultaneously maintaining robustness to heavy-tailed idiosyncratic errors.

The remainder of the paper is organized as follows. Section \ref{method} provides a concise overview of prominent conditional alpha testing procedures and introduces a robust spatial-sign-based max-type test. In Section \ref{combine}, we establish the asymptotic independence between the sum-type spatial-sign-based test and our newly proposed max-type counterpart, and then construct a Cauchy combination test based on this result. Section \ref{sec:simu} reports simulation results evaluating the finite-sample performance of the proposed test, while Section \ref{Emp} provides an empirical application to real-world financial data. Section \ref{conclusion} concludes the paper. All technical proofs are relegated to the Appendix.

\textsc{Notations.} For $k$-dimensional vector $\boldsymbol{a}$, we use the notation $\|\boldsymbol{a}\|$, $\|\boldsymbol{a}\|_{\infty}$ to denote its Euclidean norm and maximum-norm respectively. {\color{black} Additionally, let \(U(\boldsymbol{a})\) represent the spatial sign function, defined as \(U(\boldsymbol{a}) = I(\boldsymbol{a} \neq \bm{0}) \boldsymbol{a}/\|\boldsymbol{a}\|\), where \(I(\cdot)\) is the indicator function.} For an $m \times n$ matrix $\mathbf{M}=\left(m_{j \ell}\right)_{m \times n}$, the 1- and 2-norms of $\mathbf{M}$ are $\|\mathbf{M}\|_1=\max _{1 \leqslant \ell \leqslant n} \sum_{j=1}^{m}\left|m_{j \ell}\right|$ and $\|\mathbf{M}\|_2=\left\{\lambda_{\max }\left(\mathbf{M}^{\top} \mathbf{M}\right)\right\}^{1 / 2}$. The Frobenius norm of $\mathbf{M}$ is $\|\mathbf{M}\|_F=\left\{\sum_{j=1}^{m} \sum_{\ell=1}^{n} m_{j \ell}^2\right\}^{1 / 2}$.
Denote $a_n \lesssim b_n$ if there exists constant $C, a_n \leq C b_n$ and $a_n \asymp b_n$ if both $a_n \lesssim b_n$ and $b_n \lesssim a_n$ hold. For two sequences of numbers $\left\{a_n \geq 0 ; n \geq 1\right\}$ and $\left\{b_n>0 ; n \geq 1\right\}$, we write $a_n \ll b_n$ if $\lim _{n \rightarrow \infty} \frac{a_n}{b_n}=0$. {\color{black} The notations $\cp$ and $\cd$ denote convergence in probability and convergence in distribution, respectively.}  The following assumption will be imposed. Let $\psi_\vartheta(x)=\exp \left(x^\vartheta\right)-1$ be a function defined on $[0, \infty)$ for $\vartheta>0$. Then the Orlicz norm $\|\cdot\|_{\psi_\vartheta}$ of a random variable ${X}$ is defined as $\|{X}\|_{\psi_\vartheta}=\inf \left\{t>0, \mathbb{E}\left\{\psi_\vartheta(|{X}| / t)\right\} \leqslant 1\right\}$. Let $\operatorname{tr}(\cdot)$ be a trace for matrix, $\lambda_{\text {min }}(\cdot)$ and $\lambda_{\max }(\cdot)$ be the minimum and maximum eigenvalue for symmetric matrix. $\mathbf{I}_k$ represents a $k$-dimensional identity matrix, and $\operatorname{diag}\left\{v_1, v_2, \cdots, v_k\right\}$ represents the diagonal matrix with entries $\boldsymbol{v}=\left(v_1, v_2, \cdots, v_k\right)$. For $a, b \in \mathbb{R}$, we write $a \wedge b=\min \{a, b\}$.
\section{Methodology}\label{method}
While traditional asset pricing models often assume constant factor loadings for theoretical parsimony, such a static specification frequently fails to capture the evolving dynamics of financial markets. Given that asset risk exposures exhibit significant temporal shifts driven by economic cycles and structural regime changes, time-varying factor models have become essential for providing a more granular framework for asset pricing and risk management.

Within this setting, the return on asset $i$ at time $t$, denoted by $Y_{it}$, is characterized by the following conditional factor model \citep{ang2012testing}:
\begin{align}\label{mod}
Y_{it}=\alpha_{it}+\bm\beta_{it}^\top \f_t+\varepsilon_{it}=\alpha_{it}+\sum_{j=1}^p\beta_{ijt}^\top \f_{jt}+\varepsilon_{it}, i=1,\cdots,N, t=1,\cdots,T.
\end{align}
where $Y_{it}$ is the return on asset $i$ at time $t$. The vector $\bm{f}_t =\left(f_{1t }, \ldots, f_{pt }\right)^{\top} \in \mathbb{R}^p$ contains $p$ observed traded factors, and $\alpha_{it}$ represents the conditional alpha. The term $\boldsymbol{\beta}_{it} =\left(\beta_{i 1t}, \ldots, \beta_{i pt}\right)^{\top} \in \mathbb{R}^p$ denotes the vector of time-varying factor loadings, and $\varepsilon_{i t}$ is the corresponding idiosyncratic error term. The time-varying nature of factor models frequently introduces more parameters than available observations, complicating identification and estimation. To address this high-dimensional challenge, we adopt the nonparametric approach of \citet{li2011testing} and \citet{fu2025distinguishing}, assuming that the conditional alpha and factor loadings are smooth functions of time. Formally, we let $\alpha_{it} = \alpha_i(t/T)$ and $\beta_{ijt} = \beta_{ij}(t/T)$, where the time-varying coefficients $\alpha_{it}$ and $\boldsymbol{\beta}_{it}$ represent the asset's conditional alpha and beta at each time point $t$.

Evaluating whether the conditional alphas are zero across all assets $i$ and time periods $t$ constitutes a natural, albeit overly stringent, test of the pricing model. This approach is equivalent to testing the joint significance of $N \times T$ conditional alphas, a hypothesis that is frequently rejected in empirical studies. Consequently, recent literature focuses on the time-averaged conditional alphas, $T^{-1} \sum_{t=1}^T \alpha_{it}$, as a more robust measure of long-run pricing performance. Following the frameworks of \citet{lewellen2006conditional,li2011testing,ang2012testing}, we reformulate the conditional factor model to isolate the time-averaged component:
\begin{align}\label{mod_reparam}
	Y_{it} = \delta_i + \delta_i(t/T) + \sum_{j=1}^p \beta_{ij}(t/T)f_{jt} + \varepsilon_{it},
\end{align}where $\delta_i = T^{-1} \sum_{t=1}^T \alpha_{it}$ represents the long-run average pricing error for asset $i$, and $\delta_i(t/T) = \alpha_i(t/T) - \delta_i$ captures the temporal deviation from this mean. Under this decomposition, the global null hypothesis that the average conditional alphas are jointly zero across all $N$ assets is defined as:
\begin{align}\label{H0}
	H_0: \delta_i = 0 \quad \text{for all } i = 1, \ldots, N.
	\end{align}
	
To estimate the unknown time-varying parameters $\delta_i(t/T)$ and $\beta_{ij}(t/T)$, we employ a polynomial spline approach. This sieve expansion allows for a flexible approximation of the functional coefficients by projecting them onto a B-spline basis space of diverging dimensions.  Let $0 = \ell_0 < \ell_1 < \dots < \ell_n < \ell_{n+1} = 1$ denote a sequence of $n$ interior knots on the unit interval $[0,1]$. Following the literature, these knots are selected to satisfy the standard mesh condition:
\begin{equation}
	\frac{\max_{0 \le i \le n} | \ell_{i+1} - \ell_i |}{\min_{0 \le i \le n} | \ell_{i+1} - \ell_i |} \le c,
\end{equation}
for some finite positive constant $c$, where the number of knots $n = n(N, T) \to \infty$ as $(N, T) \to \infty$ \citep{su2012sieve}. For each $t$, we define its location index $i(t)$ as the unique integer satisfying $\ell_{i(t)} \le t/T < \ell_{i(t)+1}$.
On this partition consider the space of polynomial splines of order $q$ and let
$
\mathbf{B}(t/T)=(B_1(t/T),\ldots,B_L(t/T))^\top
$
be the normalized B-spline basis for that space, where $L=n+q$ \citep{de1978practical}.  For identification we work with the centered spline
basis functions
\[
\tilde B_k(t/T)=B_k(t/T)-\frac{1}{T}\sum_{s=1}^T B_k(s/T),\qquad k=1,\dots,L,
\]
and collect $\tilde{\mathbf{B}}(t/T)=(\tilde B_1(t/T),\dots,\tilde B_L(t/T))^\top$ below. The unknown time-varying coefficients are then approximated by the B-spline
functions \citep{lai1981spline}.
Specifically,
	$$
	\alpha_i(t / T) \approx \sum_{k=1}^L \lambda_{i 0 k} B_k(t / T), \quad \beta_{i j}(t / T) \approx \sum_{k=1}^L \lambda_{i j k} B_k(t / T).
	$$
	 Thus,
	\begin{align}\label{5}
		Y_{i t} & \approx \sum_{k=1}^L \lambda_{i 0 k} B_k(t / T)+\sum_{j=1}^p \sum_{k=1}^L \lambda_{i j k} B_k(t / T) f_{j t}+\varepsilon_{i t} \n\\
		& \approx \delta_i+\sum_{k=1}^L \lambda_{i 0 k}\Big\{B_k(t / T)-T^{-1} \sum_{t=1}^T B_k(t / T)\Big\}+\sum_{j=1}^p \sum_{k=1}^L \lambda_{i j k} B_k(t / T) f_{j t}+\varepsilon_{i t}
	\end{align}
Under the null hypothesis, we can estimate $\bm \lambda_i=\left(\bm\lambda_{i j}, 0 \leq j \leq p\right), \bm\lambda_{i j}=\left(\lambda_{i j 1}, \ldots, \lambda_{i j L}\right)$ by minimizing
$$
Q\left(\boldsymbol{\lambda}_i\right)=\sum_{t=1}^T\left(Y_{i t}-\bm\lambda_{i 0}^{\top} \tilde{\mathbf{B}}(t / T)-\sum_{j=1}^p \bm\lambda_{i j}^{\top} \mathbf{B}(t / T) f_{j t}\right)^2,
$$
where $\mathbf{B}(t / T)=\left(B_1(t / T), \ldots, B_L(t / T)\right) \quad$ and $\quad \tilde{\mathbf{B}}(t / T)=\left(\tilde{B}_1(t / T), \ldots, \tilde{B}_L(t / T)\right)$, $\tilde{B}_k(t / T)=B_k(t / T)-\frac{1}{T} \sum_{t=1}^T B_k(t / T) . $
 Define $$\quad \boldsymbol{Z}_t=\left\{Z_{t k}, 1 \leq k \leq(1+p) L\right\}^{\top}= \left\{\tilde{\mathbf{B}}(t / T), \boldsymbol{f}_t^{\top} \otimes \mathbf{B}(t / T)^{\top}\right\}^{\top}.$$ Thus, we can rewrite (\ref{5}) as
$$
Y_{i t} \approx \delta_i+\bm\lambda_i^{\top} \boldsymbol{Z}_t+\varepsilon_{i t} .
$$
The least-square estimator of $\boldsymbol{\lambda}_i$ is $\widehat{\boldsymbol{\lambda}}_i=\left(\boldsymbol{Z}^{\top} \boldsymbol{Z}\right)^{-1} \boldsymbol{Z}^{\top} \boldsymbol{Y}_{i\cdot}$ where $\boldsymbol{Y}_{i\cdot}=\left(Y_{i 1}, \ldots, Y_{i T}\right)^{\top}$ and $\mathbf{Z}=\left(\boldsymbol{Z}_1, \ldots, \boldsymbol{Z}_T\right)^{\top}$. So the resulting residuals are $\hat{\varepsilon}_{i t}=Y_{i t}-\widehat{\boldsymbol{\lambda}}_i \boldsymbol{Z}_t$.
As the number of interior knots is typically unknown in practice, we follow the approach of \cite{ma2014partially,ma2015varying,ma2020testing,ma2024adaptive} by employing the Bayesian information criterion (BIC) for selection. Specifically, the optimal number of knots, $n$, is determined by minimizing:
\begin{align}\label{BIC}
BIC(n) = &\log \left\{ \frac{1}{NT} \sum_{i=1}^{N} \sum_{t=1}^{T} \left( Y_{it} - \hat{\lambda}_{i0}^\top \tilde{B}(t/T) - \sum_{j=1}^{p} \hat{\lambda}_{ij}^\top B(t/T) f_{jt} \right)^2 \right\}\n\\
&\quad\quad+ \frac{\log(NT)}{NT}(p+1)(n+q).
\end{align}
 For notational convenience, we write 
$\boldsymbol{Y}_{\cdot t}=(Y_{1t}, \cdots, Y_{Nt})^\top\in\mR^N$ and $\Y=(\boldsymbol{Y}_{1\cdot}, \cdots, \boldsymbol{Y}_{N\cdot})\in\mR^{T\times N}$. And $\bmv_{i\cdot}=(\varepsilon_{i1}, \cdots, \varepsilon_{iT})^\top\in\mR^T$, 
$\bmv_{\cdot t}=(\varepsilon_{1t}, \cdots, \varepsilon_{Nt})^\top\in\mR^N$, $\bmv=(\bmv_{1\cdot}, \cdots, \bmv_{N\cdot})\in\mR^{T\times N}$, $\hat{\bmv}_{\cdot t}=(\hat{\varepsilon}_{1t}, \cdots, \hat{\varepsilon}_{Nt})^\top\in\mR^N$ and $\hat{\bmv}_{i\cdot }=(\hat{\varepsilon}_{i1}, \cdots, \hat{\varepsilon}_{iT})^\top\in\mR^T$. 
\subsection{Test procedure}
To achieve robustness against heavy-tailed innovations or potential outliers in high-dimensional settings, we adopt the spatial-sign framework following \citet{Feng2016Multivariate}. For a random sample $\{\boldsymbol{X}_i\}_{i=1}^n \subset \mathbb{R}^k$, the spatial median $\hat{\boldsymbol{\mu}}$ is obtained by minimizing $\sum_{i=1}^n (\|\boldsymbol{X}_i-\boldsymbol{\mu}\|-\|\boldsymbol{X}_i\|)$, which equivalently satisfies the following estimating equation:
\begin{equation}
	\frac{1}{n}\sum_{i=1}^n U(\boldsymbol{X}_i-\hat{\boldsymbol{\mu}})=\mathbf{0}.
\end{equation}
In our context, given the residuals $\{\hat{\bmv}_{\cdot t}\}_{t=1}^T$, we jointly estimate the location parameter $\boldsymbol{\theta}$ and a diagonal scaling matrix $\mathbf{D}$ that accounts for heteroskedasticity across dimensions. The estimators $(\hat{\boldsymbol{\theta}}, \hat{\mathbf{D}})$ are defined as the solution to the following system of M-estimation equations:
\begin{align}
	&\frac{1}{T}\sum_{t=1}^T U(\D^{-1/2}(\hat{\bmv}_{\cdot t}-\bm \theta))=0,\\
	&\frac{1}{T}\sum_{t=1}^T\diag\{ U(\D^{-1/2}(\hat{\bmv}_{\cdot t}-\bm \theta))U(\D^{-1/2}(\hat{\bmv}_{\cdot t}-\bm \theta))^\top\}=\frac{1}{N}\I_N.
\end{align}
	To solve this system, we implement an iterative algorithm as detailed in Algorithm \ref{algo1}.
	\begin{algorithm}[H]\caption{Spatial median estimator of $\hat{\bmv}$}\label{algo1}
	\begin{algorithmic}[1]
		\State Initialize $\bm{\theta}$ and $\mathbf{D}$ as the sample mean and variance of $\{\hat{\bmv}_{\cdot t}\}_{t=1}^T$;
		\State $\boldsymbol  \xi_t \gets \D^{-1/2}(\hat{\bmv}_{\cdot t}-\bm \theta)$,
		~~$t=1,\cdots,T$;
		\State $\bm \theta \gets \bm \theta+\frac{\D^{1/2}\sum_{t=1}^T U(\bm\xi_t)}{\sum_{t=1}^T ||\bm\xi_t||^{-1}}$;
		\State $\D \gets N
		\D^{1/2}\diag\{T^{-1}\sum_{t=1}^{T}U(\boldsymbol  \xi_t)U(\boldsymbol  \xi_t)^\top \}\D^{1/2}$;
		\State Repeat Steps 2-4 until convergence.
	\end{algorithmic}
	\end{algorithm}
	Based on the converged estimators $(\hat{\boldsymbol{\theta}}, \hat{\mathbf{D}})$, we define the standardized residuals as $\tilde{\bmv}_{t\cdot} = \hat{\mathbf{D}}^{-1/2}(\hat{\bmv}_{\cdot t}-\hat{\boldsymbol{\theta}})$. To construct the test statistic, we calculate the following auxiliary moment estimators:
	\begin{equation}
\hat{\varsigma}_2=\frac{1}{T}\sum_{t=1}^T\|\tilde{\bmv}_{t\cdot}\|^2,\quad
\hat{\varsigma}_{-1}=\frac{1}{T}\sum_{t=1}^T\|\tilde{\bmv}_{t\cdot}\|^{-1},\quad
\hat{\varsigma}_{1}=\frac{1}{T}\sum_{t=1}^T\|\tilde{\bmv}_{t\cdot}\|.
\end{equation} 
Let $\P_{\mathbf{Z}}=\mathbf{I}_T-\mathbf{Z}(\mathbf{Z}^\top\mathbf{Z})^{-1}\mathbf{Z}^\top$  denote the projection matrix, and define $\omega_T=\mathbf{1}_T^\top\P_{\mathbf{Z}}\mathbf{1}_T$. The adaptive scaling factor is then given by:
\begin{equation}
	\hat{\zeta}
	=\frac{N(\hat{\varsigma}_{-1})^2}{
		1-2(1-\omega_T/T)\hat{\varsigma}_{-1}\hat{\varsigma}_{1}
		+(1-\omega_T/T)\hat{\varsigma}_{2}(\hat{\varsigma}_{-1})^2 }.
	\end{equation}
	We propose the spatial-sign max-type statistic:
	\begin{equation}\label{mt}
		T_{CSM} = T\big\|\hat{\mathbf{D}}^{-1/2}\hat{\boldsymbol{\theta}}\big\|_\infty^2
		\hat{\zeta}-2\log N+\log\log N.
\end{equation}
Under the null hypothesis $H_0$ and certain regularity conditions specified in Section \ref{theot resu}, $T_{CSM}$ converges in distribution to a Type I extreme value distribution with the cumulative distribution function $ \exp\{-\pi^{-1/2}\exp(-y/2)\}$. Consequently, at a significance level $\gamma$, we reject $H_0$ if $T_{CSM} > q_\gamma$, where the critical value is given by $q_\gamma = -\log\pi - 2\log\log(1-\gamma)^{-1}$. We refer to this test as the CSM test.
 
\subsection{Related works}

To contextualize our robust framework, we briefly review existing methodologies for testing high-dimensional alphas in conditional factor models. Under the null hypothesis $H_0$, \cite{ma2020testing} proposed a sum-type test statistic to detect dense alternatives (hereinafter referred
to as HDA test):
$$J_{N T}=N^{-1} \sum_{i=1}^N\left(T^{-1 / 2} \sum_{t=1}^T \hat{\varepsilon}_{i t}\right)^2.$$
They established that $J_{N T}$ is asymptotically normally distributed within the framework of diverging factor models \citep{bai1996effect}. Let $p_{HDA}$ denote the $p$-value associated with $J_{NT}$ derived from its asymptotic normal limit. It is important to note that this theoretical validity requires the idiosyncratic error terms to possess light-tailed distributions. To address sparse alternatives, \cite{ma2024adaptive} introduced a max-type test statistic (hereinafter referred to as MNT test):
$$M_{N T}=\max _{1 \leq i \leq N} T^{-1} \widehat{\sigma}_{i i}^{-1}\left(\hat{\bmv}_{i \cdot}^{\prime} \mathbf{1}_T\right)^2,$$
where $\widehat{\sigma}_{i j}=\hat{\bmv}_{i \cdot}^{\prime} \hat{\bmv}_{j \cdot} /(T-p-1)$. Under the assumption of sub-Gaussian-type tails, they showed that $M_{NT} - 2\log(N) + \log(\log(N))$ converges in distribution to a Type I extreme value distribution. The $p$-value for the max-type test, denoted as $p_{MNT}$, is calculated based on this extreme value limiting distribution. Furthermore, leveraging the asymptotic independence between $J_{NT}$ and $M_{NT}$, \cite{ma2024adaptive} constructed an adaptive test (hereinafter referred to as Ada test) using the Cauchy combination method \citep{liu2020}. This approach aggregates $p_S$ and $p_M$ into a single $p$-value, $p_{Ada}$, defined as:
$$p_{Ada}=1-G\left[0.5 \tan \left\{\left(0.5-p_{MNT}\right) \pi\right\}+0.5 \tan \left\{\left(0.5-p_{HDA}\right) \pi\right\}\right],$$
where $G(\cdot)$ is the cumulative distribution function of the standard Cauchy distribution. We reject $H_0$ if $p_{Ada} < \gamma$ for a significance level $\gamma \in(0,1)$. While these testing procedures are effective under light-tailed errors, their size control and power deteriorate when the errors exhibit heavy-tailed behavior. These limitations motivate the need for robust testing procedures that can accommodate heavy-tailed distributions. To address this, \cite{zhao2023robust} proposed a sum-type test built on spatial signs (hereinafter referred to as CSS test): 
\begin{align}\label{CSS} 
T_{CSS}=\frac{\left(\boldsymbol{h}^{\top} \boldsymbol{h}\right)^{-1} \boldsymbol{h}^{\top}\left(U(\hat{\bmv}_{\cdot 1}), \ldots, U(\hat{\bmv}_{\cdot T})\right)^{\top}\left(U(\hat{\bmv}_{\cdot 1}), \ldots, U(\hat{\bmv}_{\cdot T})\right) \boldsymbol{h}-1}{\sqrt{\widehat{\operatorname{tr}\left(\boldsymbol{\Sigma}_{u}^2\right)}}}
\end{align}
where $ \boldsymbol{h}=\mathbf{M}_{\mathbf{Z}} \mathbf{1}_T$. It was further shown that $T_{CSS}\cd N(0,1),$
with the estimator for the trace of the second moment matrix given by:
$$
\widehat{\operatorname{tr}\left(\boldsymbol{\Sigma}_{u}^2\right)}=\frac{1}{\boldsymbol{h}^{\top} \boldsymbol{h}\left(\boldsymbol{h}^{\top} \boldsymbol{h}-1\right)} \underset{t_1, t_2=1, t_1 \neq t_2}{\sum^T\sum^T} h_{t_1}^2 h_{t_2}^2\left\{U\left(\tilde{\bmv}_{t_1}\right)^{\top} U\left(\tilde{\bmv}_{t_2}\right)\right\}^2,
$$
where $\tilde{\bmv}_t=\Y_{\cdot t}-\tilde{\mathbf{\L}} \tilde{\boldsymbol{Z}}_t$, $ \tilde{\mathbf{\L}}=\big(\tilde{\boldsymbol{\lambda}}_1, \ldots, \tilde{\boldsymbol{\lambda}}_N\big), $ and $\tilde{\boldsymbol{\lambda}}_i=\big(\tilde{\mathbf{Z}}^{\top} \tilde{\mathbf{Z}}\big)^{-1} \tilde{\mathbf{Z}}^{\top} \boldsymbol{Y}_i$. Additionally, $\tilde{\mathbf{Z}}=(\tilde{\boldsymbol{Z}}_1,\dots,\tilde{\boldsymbol{Z}}_T)^{\top}$ and $\tilde{\boldsymbol{Z}}_t=\left\{\tilde{Z}_{t k}, 1 \leq k \leq\right. (1+p) L\}^{\top}=\left\{\mathbf{B}(t / T), \boldsymbol{f}_t^{\top} \otimes \mathbf{B}(t / T)^{\top}\right\}^{\top}$.
 While the test based on $T_{CSS}$ attains good power under dense alternatives, its effectiveness drops substantially when signals are sparse. Crucially, the literature currently lacks a unified framework that simultaneously guarantees robustness to heavy tails and high power against sparse alternatives. This gap motivates our proposed approach.

 \subsection{Theoretical results}\label{theot resu}
 Now, let $\mathcal{H}_r$ denote the collection of all functions on $[0,1]$ such that the $q$-th order derivative satisfies the H\"older condition of order $m$ with $r \equiv q+m$. That is, there exists a constant $C_0 \in (0, \infty)$ such that for each $\phi \in \mathcal{H}_r$,
 $$
 \left|\phi^{(q)}\left(u_1\right)-\phi^{(q)}\left(u_2\right)\right| \leq C_0\left|u_1-u_2\right|^m
 $$
 for any $0 \leq u_1, u_2 \leq 1$. Let $\mathcal{F}_{N T, t}$ be the $\sigma$-algebra generated from $\left\{\mathbf{F},\left\{\varepsilon_{i t}, \varepsilon_{i t-1}, \ldots\right\}_{i=1}^N\right\}$, where $\quad \mathbf{F}=\left\{\bm{f}_1^{\top}, \ldots, \bm{f}_T^{\top}\right\}^{\top}$. Denote $\quad \mathbf{E}_{-t}=\big\{\varepsilon_{i 1}, \ldots,\varepsilon_{i t-1}, \varepsilon_{i t+1}, \ldots, \varepsilon_{i T}\big\}_{i=1}^N.$ 
Next, we establish the theoretical properties of the proposed statistic $T_{CSM}$.
To this end, we impose the following conditions:
\begin{itemize}
\item[(C1)] $\delta_i(\cdot) \in \mathcal{H}_r$ and $\beta_{i j}(\cdot) \in \mathcal{H}_r$ for some $r>3 / 2$.
\item[(C2)] The $p$-dimensional vector of common factors $\boldsymbol{f}_t$ is independent of the error terms $\varepsilon_{i t^{\prime}}$ for all $i=1, \ldots, N$ and $t, t^{\prime}=1, \ldots, T$. The number of factors $p$ is fixed and $\boldsymbol{f}_t^{\top} \boldsymbol{f}_t \leq K<+\infty$ for a constant $K$ and all $t=1, \ldots, T$. The matrix $T^{-1}\left(\mathbf{1}_T, \mathbf{F}\right)^{\top}\left(\mathbf{1}_T, \mathbf{F}\right)$ is positive-definite, and as $T \rightarrow \infty, T^{-1} \mathbf{1}_T^{\top} \mathbf{M}_{\mathbf{F}} \mathbf{1}_T>\tau_{\min }$ for some positive constant $\tau_{\min }$, where $\mathbf{M}_{\mathbf{F}}=\mathbf{I}_T-\mathbf{F}(\mathbf{F}^\top\mathbf{F})^{-1}\mathbf{F}^\top$.
\item[(C3)]
We consider the following model for error term:
\begin{align}\label{modelx}
	\bmv_{\cdot t}=v_t\mathbf\Gamma \boldsymbol W_t,
\end{align}
where $\boldsymbol W_t$ is a
p-dimensional random vector with independent components, $\mathbb{E}(\boldsymbol W_t)=0$, $\mathbf \Sigma=\mathbf \Gamma\mathbf\Gamma^\top$, $v_t$ is a nonnegative univariate random variable and is independent with the spatial sign of $\boldsymbol W_t$.
\item[(C4)]
(i) $W_{t, 1}, \ldots, W_{t, N}$ are i.i.d. symmetric random variables with $\mathbb{E}\left(W_{t, j}\right)=0, \mathbb{E}\left(W_{t, j}^2\right)=1$, $\left\|v_t\right\|_{\psi_{\color{black}\vartheta}} \leqslant c_1$ and $\left\|W_{t, j}\right\|_{\psi_{\color{black}\vartheta}} \leqslant c_0$ with some constants $c_0,c_1>0$ and $1 \leqslant {\color{black}\vartheta} \leqslant 2$. (ii) Let $\D$ is the diagonal matrix of $\bms$ and $r_t=||\D^{-1/2}\bmv_{\cdot t}||$. The moments $\zeta_k=\mathbb{E}\left(r_t^{-k}\right)$ for $k=1,2,3,4$ exist for large enough $N$. In addition, there exist two positive constants $\underline{b}$ and $\bar{B}$ such that $\underline{b} \leqslant \lim \sup_N \mathbb{E}\left(r_t / \sqrt{N}\right)^{-k} \leqslant \bar{B}$ for $k=1,2,3,4$.
\item[(C5)] (i) The shape matrix $\R=\mathbf D^{-1/2}{\bf \Gamma\Gamma}^\top \mathbf D^{-1/2}=\left(\sigma_{j \ell}\right)_{N \times N}$  satisfies $\max _{j=1,\cdots,N}\sum_{\ell=1}^N\left|\sigma_{j \ell}\right| \leqslant a_0(N)$. We assume that $a_0(N)\asymp N^{1-\delta}$, ${\color{black}0<\delta\leq1/2}$. Let $TL^{-2r}=o(1)$, $L^3T^{-1}=o(1)$, $LN^{-1/2}=o(1)$, $\log N=o(T^{1/5})$ and $L\log (NT)=o(N^{1/3 \wedge \delta})$. In addition, $\lim\inf_{N\rightarrow \infty}\min_{j=1,2,\cdots,N}{\color{black}d}_j>\underline{d}$ for some constant $\underline d>0$, where $\mathbf D=\operatorname{diag}\{d_1^2,d_2^2,\cdots,d_N^2\}$. (ii)   For some $\varrho \in(0,1)$, assume $|\sigma_{ij}|\leq \varrho$ for all $1\leq i<j\leq N$ and $N
    \geq 2$. Suppose $\left\{\delta_N ; N \geq 1\right\}$ and $\left\{\kappa_N ; N \geq 1\right\}$ are positive constants with $\delta_N=o(1 / \log N)$ and $\kappa=\kappa_N \rightarrow 0$ as $N \rightarrow \infty$. For $1 \leq i \leq N$, define $B_{N, i}=\left\{1 \leq j \leq N ;\left|\sigma_{i j}\right| \geq \delta_N\right\}$ and $C_N=\left\{1 \leq i \leq N ;\left|B_{N, i}\right| \geq N^\kappa\right\}$. We assume that $\left|C_N\right| / N \rightarrow 0$ as $N \rightarrow \infty$.
\end{itemize}
\begin{remark}{}
	Following \cite{he1996bivariate}, Condition (C1) provides the necessary smoothness regularity for the unknown functions, a common prerequisite in nonparametric estimation settings. Condition (C1) is identical to Assumption (A1) in \cite{ma2020testing}, which is extensively used in high-dimensional conditional alpha testing procedures, such as those found in \cite{zhao2023robust,ma2024adaptive}. Condition (C2) Assumption (A2) requires that the factors and idiosyncratic errors are mutually independent and that the factor energy, $\bm{f}_t^\top \bm{f}_t$, is uniformly bounded. This condition is equivalent to Assumption (A1) in \cite{fan2015power,lan2018testing,pesaran2023testing}. Condition (C3) specifies the error structure (\ref{modelx}), which is analogous to model (3) in \cite{cheng2023}. This setup is sufficiently flexible to encompass a wide array of multivariate distribution families and models, most notably the family of elliptical distributions \citep{hallin2006semiparametrically,Oja2010Multivariate,fang2018symmetric} and the independent components model \citep{nordhausen2009signed,ilmonen2011semiparametrically,yao2015sample}. Regarding Condition (C4), it is consistent with Conditions (C.1) and (C.2) of \cite{cheng2023}. Under the symmetry assumption invoked here, the spatial median of the residuals $\{\hat{\bmv}_{\cdot t}\}_{t=1}^T$ is identified as $T^{-1}\omega_T \bm \delta$. Furthermore, part (ii) of Condition (C4) generalizes Assumption 1 in \cite{zou2014multivariate} by stipulating that $\zeta_k \asymp N^{-k/2}$; for a more exhaustive discussion of these properties, we refer readers to \cite{cheng2023}. Finally, Condition (C5) restricts the degree of cross-sectional correlation among the variables. It is easily verified that common structures, such as AR(1) processes or banded matrices with a fixed bandwidth, satisfy this requirement.
\end{remark}
Theorem \ref{th1} below states the limiting distribution of the spatial-median–based test statistic under the null hypothesis.
\begin{theorem}\label{th1}
Under the null hypothesis and Conditions (C1)-(C5), we have
\begin{align*}
P\left(T||\hat{\D}^{-1/2}\hat{\bm \theta}||^2_\infty\zeta-2\log N+\log\log N\le y\right)\to \exp \left\{-\frac{1}{\sqrt{\pi}}\exp(-y/2)\right\}
\end{align*}
where { $\zeta=N\{E(r_t^{-1})\}^2/\eta_{\omega}$, $\eta_{\omega}=1-2\eta E(r_t^{-1})E(r_t)+\eta E(r_s^{2})\{E(r_t^{-1})\}^{2}$ and $T^{-1}\omega_{T}\cp \omega=1-\eta$.}
\end{theorem}
Theorem \ref{th1} follows from the extreme-value behavior of the spatial median $\hat{\bm\theta}$. Owing to the spatial median's robustness, the proposed statistic yields reliable inference even when idiosyncratic errors are heavy-tailed. Practical implementation requires a consistent estimator for the nuisance scalar $\zeta$. The next proposition establishes the validity of our plug-in estimator.
\begin{prop}\label{pro1}
Under Conditions (C1)-(C4), we have $\hat{\zeta}/\zeta\cp 1$.
\end{prop}
By Slutsky's theorem, Theorem \ref{th1} together with Proposition \ref{pro1} implies that the CSM statistic $T_{CSM}$ has the asymptotic distribution
\begin{align*}
P\left(T_{CSM}\le y\right)\to \exp \left\{-\frac{1}{\sqrt{\pi}}\exp(-y/2)\right\}\doteq G(y),
\end{align*}
and the corresponding $p$-value is $p_{CSM}=1-G(T_{CSM})$.
Furthermore, we examine the power properties of the test statistic $T_{CSM}$ under local alternative hypotheses. The following theorem establishes the asymptotic consistency of the test statistic $T_{CSM}$.
\begin{theorem}
Under Conditions (C1)-(C4), for some sufficiently large constant $C>0$,
if $\|\bm \delta\|_{\infty} \geq C\sqrt{\log N/T}${,  $\|\bm \delta\|^2 =O (NT^{-1} L )$ and $\lambda_{\max}(\mathbf{R})=o(NL^{-1}(\log N)^{-1})$}, we have the power of $T_{CSM}$ goes to one.
\end{theorem}
This result demonstrates that under the specified conditions, the CSM test is asymptotically powerful against alternatives in which the maximum component of $\bm\delta$ exceeds the threshold $C\sqrt{\log N/T}$.
\section{Robust Cauchy combination test}\label{combine}

In practice, it is often difficult to ascertain whether the alternative hypothesis is dense or sparse. Therefore, there is a need to develop a new test procedure that is robust to the sparsity of the alternative hypothesis. \cite{ma2024adaptive} proposed the Ada test that combines the p-values from the HDA test and the MNT test. However, both the $J_{NT}$ and $M_{NT}$ tests lack robustness when handling heavy-tailed distributions. As such, there is a need to develop a new test procedure that is not only robust to the sparsity of the alternative hypothesis but also to heavy-tailed distributions. Hence, we suggest combining the corresponding $p$-values of CSM and CSS tests by using truncated Cauchy combination method \citep{liu2020}, to wit,
\begin{align*}
	p_{CC}&=1-F[0.5\tan\{(0.5-p_{CSS})\pi\}I(p_{CSS}<0.5)+0.5\tan\{(0.5-p_{CSM})\pi I(p_{CSM}<0.5)\}]
\end{align*}
where $F(\cdot)$ is the CDF of the standard Cauchy distribution. If the final $p$-value is less than some pre-specified significant level $\gamma\in(0,1)$, then we reject $H_0$. We refer to this test procedure CC test hereafter.

To establish the asymptotic results of CC test, we first analyze the relationship between the spatial-sign-based max-type test (\ref{mt}) and the sum-type test (\ref{CSS}).
 We impose the following condition:
\begin{itemize}
	 \item [(C6)] The error vectors $\bmv_{\cdot 1}, \ldots, \bmv_{\cdot T}$ are i.i.d. from the $N$-variate mean zero elliptical distribution with probability density function:
		$$
		\operatorname{det}(\Xi)^{-1 / 2} g\left(\left\|\Xi^{-1 / 2} \bmv\right\|\right), \bmv \in \mathcal{R}^N.
		$$
	\item [(C7)] Assume that $\operatorname{tr}\left(\boldsymbol{\Omega}^4\right)=o\left(\operatorname{tr}^2\left(\boldsymbol{\Omega}^2\right)\right)$ and $\operatorname{tr}^4(\boldsymbol{\Omega})/\operatorname{tr}^2\left(\boldsymbol{\Omega}^2\right) \exp \left\{-\operatorname{tr}^2(\boldsymbol{\Omega})/\{128 N \lambda_{\max }^2(\boldsymbol{\Omega})\}\right\} \rightarrow 0$ where $\boldsymbol{\Omega}=\operatorname{Cov}(\bmv_{\cdot t}) \doteq(\varsigma_{i j})_{1 \leq i, j \leq N}$ and $\lambda_{\text {max }}(\boldsymbol{\Omega})$ is the largest eigenvalue of $\boldsymbol{\Omega}$. Additionally, $T L^{-2 r} N\{\operatorname{tr}(\boldsymbol{\Omega}^2)\}^{-1 / 2} \rightarrow 0$, $\{\operatorname{tr}(\boldsymbol{\Omega}^2)\}^{-1 / 2} \max _i \sum_{j=1}^N|\varsigma_{i j}| \rightarrow 0$ and $$T^{-1+\varrho} N^{1+\varrho} L\{\operatorname{tr}(\boldsymbol{\Omega}^2)\}^{-1 / 2}=O(1)$$ for an arbitrarily small $\varrho>0$.
\end{itemize}
	 Note that Conditions (C6)-(C7) are identical to Assumptions (A1) and (A2) in \cite{zhao2023robust}.
\begin{theorem}\label{thm3}
Under Conditions (C1)-(C7) and $\log N=o({ T^{1/10}})$, we have $T_{CSS}$ is asymptotically independent with $T_{CSM}$ under the null hypothesis, i.e.
\begin{align*}
P\left(T_{CSS}\le x, T_{CSM}\le y\right)\to \Phi(x)G(y),
\end{align*}
where  $\Phi(x)$  denotes the cumulative distribution function of a standard normal random variable.
\end{theorem}

To analyze the power performance of the new proposed CC test, we also demonstrate that $T_{CSS}$ are also asymptotically independent with $T_{CSM}$ under some special alternatives.
{\color{black} Then, we consider the relationship between $T_{S S}$ and $T_{SM }$ under the following local alternative hypothesis:
\begin{align}\label{H_1_comb}
H_1:\boldsymbol{\delta}^\top\boldsymbol{\delta}=O\left(\varkappa_1^{-2}  T^{-1}\sqrt{2\operatorname{tr}\left(\boldsymbol{\Sigma}_{u}^2\right)} \right),\boldsymbol{\delta}^\top\mathbf{\Omega}\boldsymbol{\delta}=o\left(\varkappa_1^{-2} N^{-1} T^{-1} \operatorname{tr}(\mathbf{\Omega}^2) \right),
\end{align}
where $\boldsymbol{\Sigma}_{u} = E(U(\bmv_{\cdot t})U(\bmv_{\cdot t})^{\top})$ and $\varkappa_1=E\left(\left\|\bmv_{\cdot t}\right\|^{-1}\right)$.
 The following theorem establishes the asymptotic independence between $T_{CSM}$ and $T_{CSS}$ under this special alternative hypothesis.}
\begin{theorem}\label{thm4}
Under Conditions (C1)-(C7), $\log N=o({ T^{1/10}})$, and the alternative hypothesis \eqref{H_1_comb}, we have
\begin{align*}
P\left(T_{CSS}\le x, T_{CSM}\le y\right)\to P\left(T_{CSS}\le x\right)P\left(T_{CSM}\le y\right).
\end{align*}
\end{theorem}
According to \cite{li2023}, the Cauchy combination-based test has more power than the test based on the minimum of $p_{CSM}$ and $p_{CSS}$, which is also known as the minimal p-value combination. This is represented as $\beta_{M\wedge S, \gamma}=P(\min\{p_{CSM}, p_{CSS}\}\leq 1-\sqrt{1-\gamma})$ {\color{black}for some pre-specified significant level $\gamma\in(0,1)$}.

It is clear that:
\begin{align}\label{power_H1}
\beta_{M\wedge S, \gamma} &\ge P(\min\{{p}_{CSM},{p}_{CSS}\}\leq \gamma/2)\nonumber\\
&= \beta_{CSM,\gamma/2}+\beta_{CSS,\gamma/2}-P({p}_{CSM}\leq \gamma/2, {p}_{CSS}\leq \gamma/2)\nonumber\\
&\ge \max\{\beta_{CSM,\gamma/2},\beta_{CSS,\gamma/2}\},
\end{align}
where {\color{black}$\beta_{CC, \gamma}=P({p}_{CC}<\gamma)$, $\beta_{CSM, \gamma}=P({p}_{CSM}<\gamma)$ and $\beta_{CSS, \gamma}=P({p}_{CSS}<\gamma)$.}
On the other hand, under the local alternative hypothesis (\ref{H_1_comb}), we have:
\begin{align}\label{power_H1np}
{\color{black}\beta_{CC, \gamma}} \ge\beta_{M\wedge S, \gamma} \ge \beta_{CSM,\gamma/2}+\beta_{CSS,\gamma/2}-\beta_{CSM,\gamma/2}\beta_{CSS,\gamma/2}+o(1),
\end{align}
which is due to the asymptotic independence implied by Theorem \ref{thm4}.

For a small $\gamma$, the difference between $\beta_{CSM,\gamma}$ and $\beta_{CSM,\gamma/2}$ is small, and the same applies to $\beta_{CSS,\gamma}$. Therefore, according to equations \eqref{power_H1} and \eqref{power_H1np}, the power of the adaptive test is at least as large as, or even significantly larger than, that of either the max-type or sum-type test. For a detailed comparison of the power performance of the max-type, sum-type, and Cauchy combination-based tests under varying conditions of sparsity and signal strength, please refer to Table 1 in \cite{ma2024testing}.

\section{Simulation}\label{sec:simu}
To evaluate the finite-sample performance of the proposed CSM and CC tests, we design three simulation scenarios inspired by \citep{ma2020testing,ma2024adaptive}. These scenarios are calibrated to replicate the stylized facts of the U.S. equity market.  For comparison, we benchmark our methods against several existing procedures, specifically the HDA \citep{ma2020testing}, MNT \citep{ma2024adaptive}, and Ada \citep{ma2024adaptive} tests.
\begin{example}\label{ex1}
	 Following \cite{li2011testing}, we generate returns from a conditional CAPM with time-varying intercepts:
	$$
	Y_{i t}=\alpha_{i t}+\beta_{i t} f_t+\varepsilon_{i t}, \quad i=1, \ldots, N, \quad t=1, \ldots, T,
	$$
	where $f_t$ denotes the market excess return. To mimic U.S. market dynamics, we specify $f_t$ as an $\operatorname{AR}(1)-\operatorname{GARCH}(1,1)$ process:
	$$
	f_t-0.34=0.05\left(f_{t-1}-0.34\right)+h_t^{1 / 2} \varphi_t,
	$$
	where $\varphi_t$ follows a standard normal distribution, and the conditional variance $h_t$ evolves according to:
	$$
	h_t=0.32+0.67 h_{t-1}+0.13 h_{t-1} \varphi_{t-1}^2,
	$$
The coefficients are calibrated by fitting the model to historical U.S. stock market data. We specify the conditional factor loadings \citep{su2017time} as a deterministic smooth function of time: $\beta_{it} = G(10t/T, 2, 2)$, where $G(z, \kappa_1, \kappa_2) = [1 + \exp\{-\kappa_1(z - \kappa_2)\}]^{-1}$ is the logistic function with tuning parameter $\kappa_1$ and location parameter $\kappa_2$.
	\end{example}
		\begin{example}\label{ex2}
		We consider a conditional Fama-French three-factor model to examine the impact of multiple factors on test performance:
			$$
			Y_{i t}=\alpha_{i t}+\sum_{j=1}^3 \beta_{i j t} f_{j t}+\varepsilon_{i t}(i=1, \ldots, N, t=1, \ldots, T),
			$$
			where $f_{1t}, f_{2t}$, and $f_{3t}$ represent the Market, SMB (small [size] minus big), and HML (high [value] minus low) factors, respectively. These factors are simulated as independent $\operatorname{AR}(1)-\operatorname{GARCH}(1,1)$ processes:
			$$
			\begin{aligned}
				& \text { Market factor: } f_{1 t}-0.34=0.05\left(f_{1 t-1}-0.34\right)+h_{1 t}^{1 / 2} \varphi_{1 t}, \\
				& \text { SMB factor: } f_{2 t}-0.04=0.07\left(f_{2 t-1}-0.04\right)+h_{2 t}^{1 / 2} \varphi_{2 t}, \\
				& \text { HML factor: } f_{3 t}-0.06=0.04\left(f_{3 t-1}-0.06\right)+h_{3 t}^{1 / 2} \varphi_{3 t},
			\end{aligned}
			$$
			where $\varphi_{j t}(j=1,2,3)$ are simulated from a standard normal distribution, $h_{j t}(j=1,2$, and 3$)$ are generated according to the following processes, 
			$$
			\begin{aligned}
				& \text { Market factor: } h_{1 t}=0.32+0.67 h_{1 t-1}+0.13 h_{1 t-1} \varphi_{1 t-1}^2, \\
				& \text { SMB: } h_{2 t}=0.33+0.51 h_{2 t-1}+0.03 h_{2 t-1} \varphi_{2 t-1}^2, \\
				& \text { HML: } h_{3 t}=0.26+0.72 h_{3 t-1}+0.05 h_{3 t-1} \varphi_{3 t-1}^2,
			\end{aligned}
			$$
			respectively.
			To evaluate the robustness of our tests against stochastic factor loadings, we model $\beta_{ijt}$ as a function of an unobservable state variable $z_t$: $\beta_{ijt} = a_j + b_j z_t$. The state variable $z_t$ follows an $\text{AR}(1)\text{-ARCH}(1)$ process: $z_t = 0.5 z_{t-1} + \sigma_t \epsilon_t$, where $\sigma_t^2 = 0.1 + 0.3\sigma_{t-1}^2$ and $\epsilon_t \sim \text{iid } N(0,1)$. The parameters are set to $(a_1, b_1)=(0.8, 0.3)$, $(a_2, b_2)=(0.5, 0.1)$, and $(a_3, b_3)=(0.6, 0.2)$.
	\end{example}
	\begin{example}\label{ex3}
	In this example, we retain the factor processes $f_{jt}$ from Example 2, but set the conditional factor loadings to be deterministic smooth functions of $t/T$:
	$$\beta_{ijt} = a_j G(10t/T, 2, 2) + b_j, \quad j=1,2,3,$$
	where the parameters are $(a_1, b_1)=(0.5, 0.5)$, $(a_2, b_2)=(0.1, 0.5)$, and $(a_3, b_3)=(0.2, 0.5)$. 
	\end{example}
	
Under the null hypothesis $H_0: \alpha_{it} = 0$ for all $i, t$, we simulate the data generating process starting from $t = -49$. Following \cite{pesaran2023testing}, we initialize the recursions at $t = -50$ by setting $f_{-50} = z_{-50} = 0$ and $h_{-50} = \sigma^2_{-50} = 1$ (with an analogous setup applied to the three-factor specification). To eliminate initialization bias, the first 50 observations are treated as a burn-in sample, and only the data for $t = 1, \dots, T$ are retained for the final experiments. The errors are generated from four scenarios with $\bms=(0.5^{|i-j|})_{1\le i,j\le N}$:
\begin{itemize}
\item[(I)] Multivariate normal distribution. $\bm \varepsilon_{\cdot t}\sim N(\bm0,\bms)$.
\item[(II)] Multivariate $t$-distribution.   $\bm \varepsilon_{\cdot t}$ are drawn from a multivariate $t$-distribution with $3$ degrees of freedom, zero mean, and scatter matrix $\bms$.
\item[(III)] Multivariate mixture normal distribution. $\bm \varepsilon_{\cdot t}$'s are generated from standardized  $[\kappa
N(\bm 0,\bms)+(1-\kappa)N(\bm 0,9\bms)]$. $\kappa$ is chosen to be 0.9.
\item[(IV)] Independent component model. $\bm \varepsilon_{\cdot t}=\bms^{1/2}\bm\epsilon_{\cdot t}$, $\bm\epsilon_{\cdot t}=(\epsilon_{t1},\cdots,\epsilon_{tN})^\top$ where $\epsilon_{ti}, i=1,\cdots,N$ are all independent and identical distributed as $t(3)/\sqrt{3}$.
\end{itemize}

In this section, we evaluate the empirical sizes of the HDA, MNT, Ada, CSS, CSM, and CC tests using 1,000 independent Monte Carlo replications. Tables \ref{t1}-\ref{t2} report the results for sample sizes $T \in \{350, 400\}$ and dimensions $N \in \{200, 400, 600\}$ at the 5\% nominal level. Overall, the empirical sizes are well-calibrated across all considered data-generating processes. Moreover, the number of
interior knots $n$ is determined by the BIC criterion and the order
of B-splines is set as 3.
 While the MNT and Ada tests appear slightly conservative in certain configurations, such as the ICM cases, none of the methods suffer from significant size distortion. These findings indicate that the proposed tests effectively control the Type I error rate even when the dimensionality is large relative to the sample size.

\begin{table}[!h]
	\begin{center}
		\caption{Sizes of tests with $T=350$.}\label{t1}
\scalebox{0.8}{
	\begin{tabular}{ccllllllcllllllclllllllllllll}\hline\hline
	       &     & \multicolumn{6}{c}{Multivariate normal distribution}         &  & \multicolumn{6}{c}{Multivariate $t$-distribution} \\ \cline{3-8} \cline{10-15} 
	     Example & N   & HDA       & MNT      & Ada     & CSS     & CSM     & CC      &  & HDA    & MNT    & Ada    & CSS    & CSM   & CC    \\
	 	  1 & 200 & 0.024 & 0.024 & 0.029 & 0.045 & 0.061 & 0.063 &  & 0.014 & 0.017 & 0.015 & 0.046 & 0.061 & 0.066 \\
	 	  1 & 400 & 0.016 & 0.016 & 0.011 & 0.042 & 0.053 & 0.051 &  & 0.004 & 0.014 & 0.008 & 0.041 & 0.053 & 0.050 \\
	 	  1 & 600 & 0.014 & 0.023 & 0.015 & 0.043 & 0.060 & 0.056 &  & 0.004 & 0.016 & 0.008 & 0.048 & 0.052 & 0.056 \\\\
	 	  2 & 200 & 0.023 & 0.016 & 0.019 & 0.045 & 0.080 & 0.075 &  & 0.016 & 0.010 & 0.011 & 0.054 & 0.072 & 0.061 \\
	 	  2 & 400 & 0.015 & 0.018 & 0.019 & 0.058 & 0.077 & 0.069 &  & 0.010 & 0.016 & 0.016 & 0.064 & 0.085 & 0.082 \\
	 	  2 & 600 & 0.020 & 0.009 & 0.016 & 0.064 & 0.071 & 0.077 &  & 0.004 & 0.016 & 0.009 & 0.047 & 0.079 & 0.065 \\\\
	 	  3 & 200 & 0.021 & 0.012 & 0.014 & 0.043 & 0.071 & 0.062 &  & 0.010 & 0.018 & 0.010 & 0.038 & 0.063 & 0.055 \\
	 	  3 & 400 & 0.017 & 0.018 & 0.019 & 0.054 & 0.086 & 0.076 &  & 0.010 & 0.014 & 0.014 & 0.039 & 0.070 & 0.065 \\
	 	  3 & 600 & 0.009 & 0.017 & 0.009 & 0.055 & 0.078 & 0.062 &  & 0.003 & 0.016 & 0.009 & 0.036 & 0.074 & 0.070\\
\\
	         &     & \multicolumn{6}{c}{Multivariate mixture normal distribution} &  & \multicolumn{6}{c}{Independent component model} \\
	         \cline{3-8} \cline{10-15} \\  
	 &     & HDA       & MNT      & Ada     & CSS     & CSM     & CC      &  & HDA    & MNT    & Ada    & CSS    & CSM   & CC    \\
	  1 & 200 & 0.020 & 0.013 & 0.020 & 0.043 & 0.052 & 0.055 &  & 0.017 & 0.010 & 0.015 & 0.042 & 0.039 & 0.042 \\
	  1 & 400 & 0.018 & 0.012 & 0.014 & 0.057 & 0.061 & 0.057 &  & 0.019 & 0.012 & 0.017 & 0.062 & 0.051 & 0.065 \\
	  1 & 600 & 0.012 & 0.014 & 0.014 & 0.041 & 0.053 & 0.051 &  & 0.017 & 0.014 & 0.012 & 0.050 & 0.043 & 0.051 \\\\
	  2 & 200 & 0.021 & 0.026 & 0.026 & 0.045 & 0.077 & 0.070 &  & 0.016 & 0.009 & 0.016 & 0.049 & 0.061 & 0.057 \\
	  2 & 400 & 0.021 & 0.016 & 0.015 & 0.062 & 0.078 & 0.079 &  & 0.018 & 0.007 & 0.021 & 0.060 & 0.049 & 0.064 \\
	  2 & 600 & 0.011 & 0.011 & 0.009 & 0.042 & 0.068 & 0.063 &  & 0.019 & 0.010 & 0.016 & 0.057 & 0.056 & 0.067 \\\\
	  3 & 200 & 0.019 & 0.020 & 0.026 & 0.047 & 0.065 & 0.065 &  & 0.018 & 0.009 & 0.012 & 0.046 & 0.049 & 0.052 \\
	  3 & 400 & 0.015 & 0.014 & 0.013 & 0.050 & 0.062 & 0.060 &  & 0.013 & 0.006 & 0.011 & 0.061 & 0.045 & 0.061 \\
	  3 & 600 & 0.015 & 0.015 & 0.011 & 0.043 & 0.063 & 0.058 &  & 0.005 & 0.013 & 0.005 & 0.050 & 0.050 & 0.060 \\
	 \hline\hline               
	\end{tabular}}
\end{center}
\end{table}

\begin{table}[!h]
	\begin{center}
		\caption{Sizes of tests with $T=400$.}\label{t2}
		\scalebox{0.8}{
			\begin{tabular}{ccllllllcllllllclllllllllllll}\hline\hline
				&     & \multicolumn{6}{c}{Multivariate normal distribution}         &  & \multicolumn{6}{c}{Multivariate $t$-distribution} \\ \cline{3-8} \cline{10-15} 
				Example & N   & HDA       & MNT      & Ada     & CSS     & CSM     & CC      &  & HDA    & MNT    & Ada    & CSS    & CSM   & CC    \\
				1 & 200 & 0.034 & 0.025 & 0.035 & 0.054 & 0.052 & 0.056 &  & 0.019 & 0.020 & 0.018 & 0.051 & 0.049 & 0.056 \\
				1 & 400 & 0.022 & 0.018 & 0.022 & 0.053 & 0.054 & 0.064 &  & 0.012 & 0.022 & 0.012 & 0.049 & 0.054 & 0.056 \\
				1 & 600 & 0.016 & 0.019 & 0.017 & 0.048 & 0.063 & 0.060 &  & 0.002 & 0.021 & 0.012 & 0.037 & 0.064 & 0.056 \\\\
				2 & 200 & 0.022 & 0.016 & 0.022 & 0.057 & 0.071 & 0.072 &  & 0.014 & 0.016 & 0.014 & 0.050 & 0.055 & 0.054 \\
				2 & 400 & 0.018 & 0.012 & 0.011 & 0.051 & 0.071 & 0.062 &  & 0.009 & 0.015 & 0.014 & 0.056 & 0.068 & 0.069 \\
				2 & 600 & 0.018 & 0.020 & 0.019 & 0.044 & 0.065 & 0.069 &  & 0.007 & 0.026 & 0.012 & 0.064 & 0.087 & 0.082 \\\\
				3 & 200 & 0.024 & 0.021 & 0.028 & 0.054 & 0.064 & 0.067 &  & 0.016 & 0.020 & 0.017 & 0.038 & 0.072 & 0.067 \\
				3 & 400 & 0.017 & 0.022 & 0.023 & 0.053 & 0.071 & 0.070 &  & 0.005 & 0.015 & 0.009 & 0.051 & 0.072 & 0.070 \\
				3 & 600 & 0.008 & 0.020 & 0.010 & 0.056 & 0.078 & 0.076 &  & 0.003 & 0.021 & 0.013 & 0.036 & 0.070 & 0.063	\\
				&     & \multicolumn{6}{c}{Multivariate mixture normal distribution} &  & \multicolumn{6}{c}{Independent component model} \\
				\cline{3-8} \cline{10-15} 
				&     & HDA       & MNT      & Ada     & CSS     & CSM     & CC      &  & HDA    & MNT    & Ada    & CSS    & CSM   & CC    \\
		1 & 200 & 0.022 & 0.015 & 0.018 & 0.043 & 0.052 & 0.057 &  & 0.020 & 0.009 & 0.014 & 0.041 & 0.034 & 0.046 \\
		1 & 400 & 0.019 & 0.012 & 0.018 & 0.046 & 0.056 & 0.058 &  & 0.020 & 0.014 & 0.016 & 0.052 & 0.046 & 0.063 \\
		1 & 600 & 0.011 & 0.013 & 0.013 & 0.049 & 0.053 & 0.050 &  & 0.015 & 0.017 & 0.007 & 0.058 & 0.053 & 0.057 \\\\
		2 & 200 & 0.021 & 0.014 & 0.022 & 0.041 & 0.074 & 0.073 &  & 0.026 & 0.011 & 0.022 & 0.061 & 0.056 & 0.066 \\
		2 & 400 & 0.021 & 0.013 & 0.020 & 0.057 & 0.060 & 0.068 &  & 0.016 & 0.011 & 0.015 & 0.063 & 0.046 & 0.065 \\
		2 & 600 & 0.011 & 0.016 & 0.012 & 0.055 & 0.076 & 0.081 &  & 0.017 & 0.010 & 0.016 & 0.063 & 0.045 & 0.064 \\\\
		3 & 200 & 0.018 & 0.023 & 0.021 & 0.052 & 0.073 & 0.070 &  & 0.020 & 0.013 & 0.016 & 0.052 & 0.051 & 0.056 \\
		3 & 400 & 0.010 & 0.015 & 0.014 & 0.042 & 0.069 & 0.064 &  & 0.014 & 0.015 & 0.013 & 0.058 & 0.063 & 0.069 \\
		3 & 600 & 0.007 & 0.016 & 0.014 & 0.047 & 0.067 & 0.064 &  & 0.008 & 0.020 & 0.014 & 0.048 & 0.084 & 0.067\\
				\hline\hline               
		\end{tabular}}
	\end{center}
\end{table}

\begin{figure}[!ht]
\centering
{\includegraphics[width=1\textwidth]{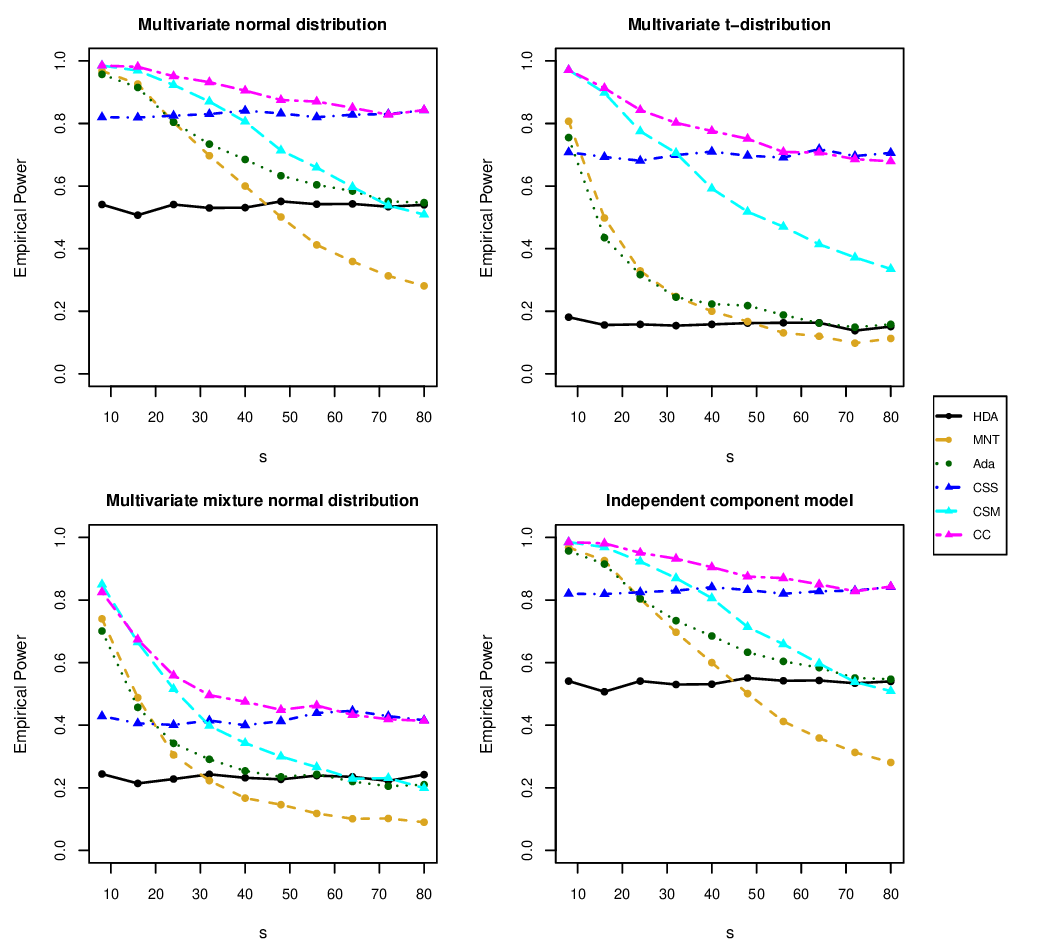}}
\caption{Power performance against sparsity level $s$ for Example \ref{ex1}. \label{powerex1}}
\end{figure}

\begin{figure}[!ht]
	\centering
	{\includegraphics[width=1\textwidth]{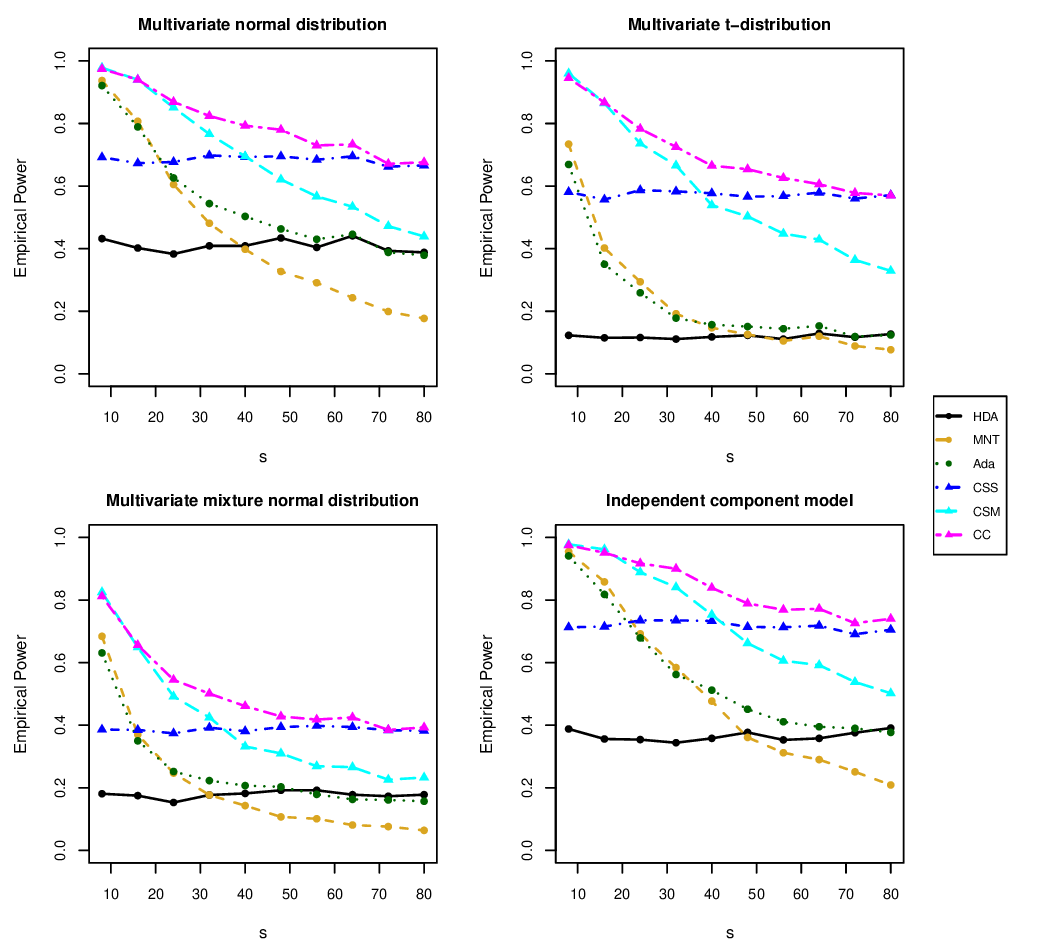}}
		\caption{Power performance against sparsity level $s$ for Example \ref{ex2}. \label{powerex2}}
\end{figure}

\begin{figure}[!ht]
	\centering
	{\includegraphics[width=1\textwidth]{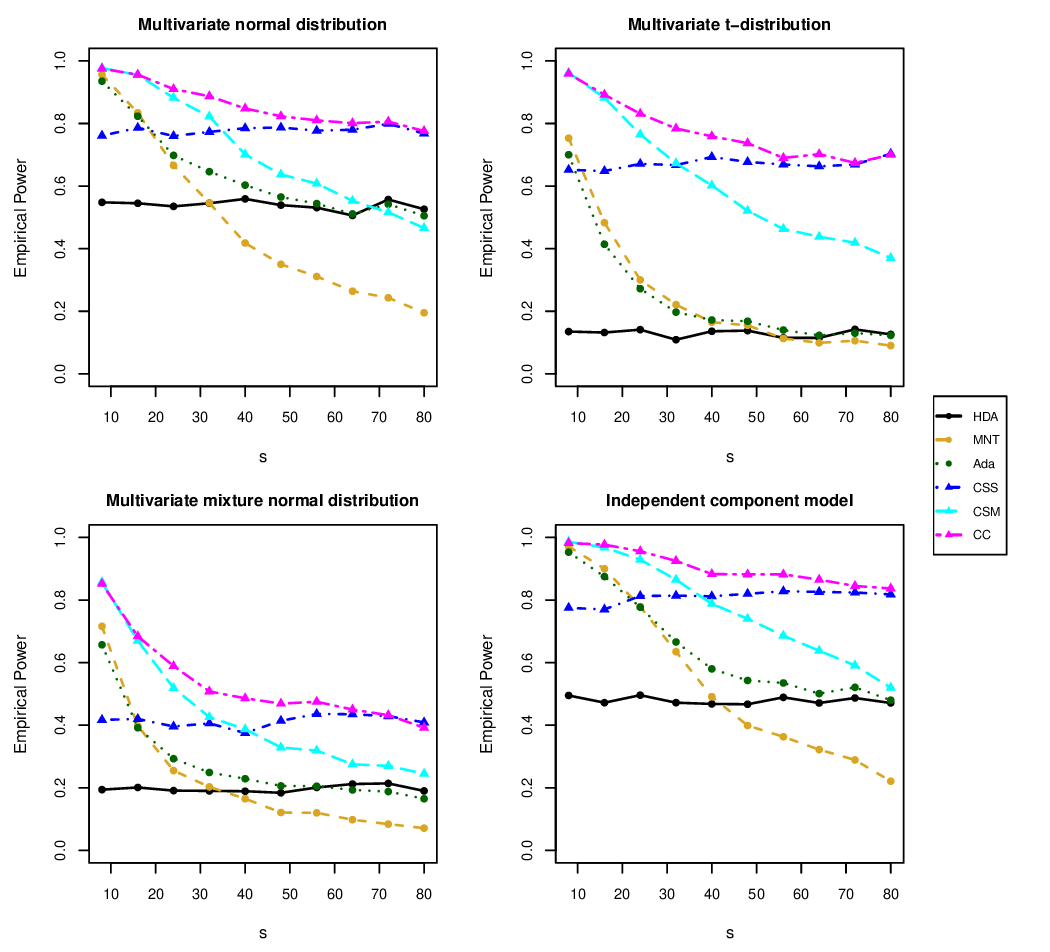}}
		\caption{Power performance against sparsity level $s$ for Example \ref{ex3}. \label{powerex3}}
\end{figure}

\begin{figure}[!ht]
	\centering
	{\includegraphics[width=1\textwidth]{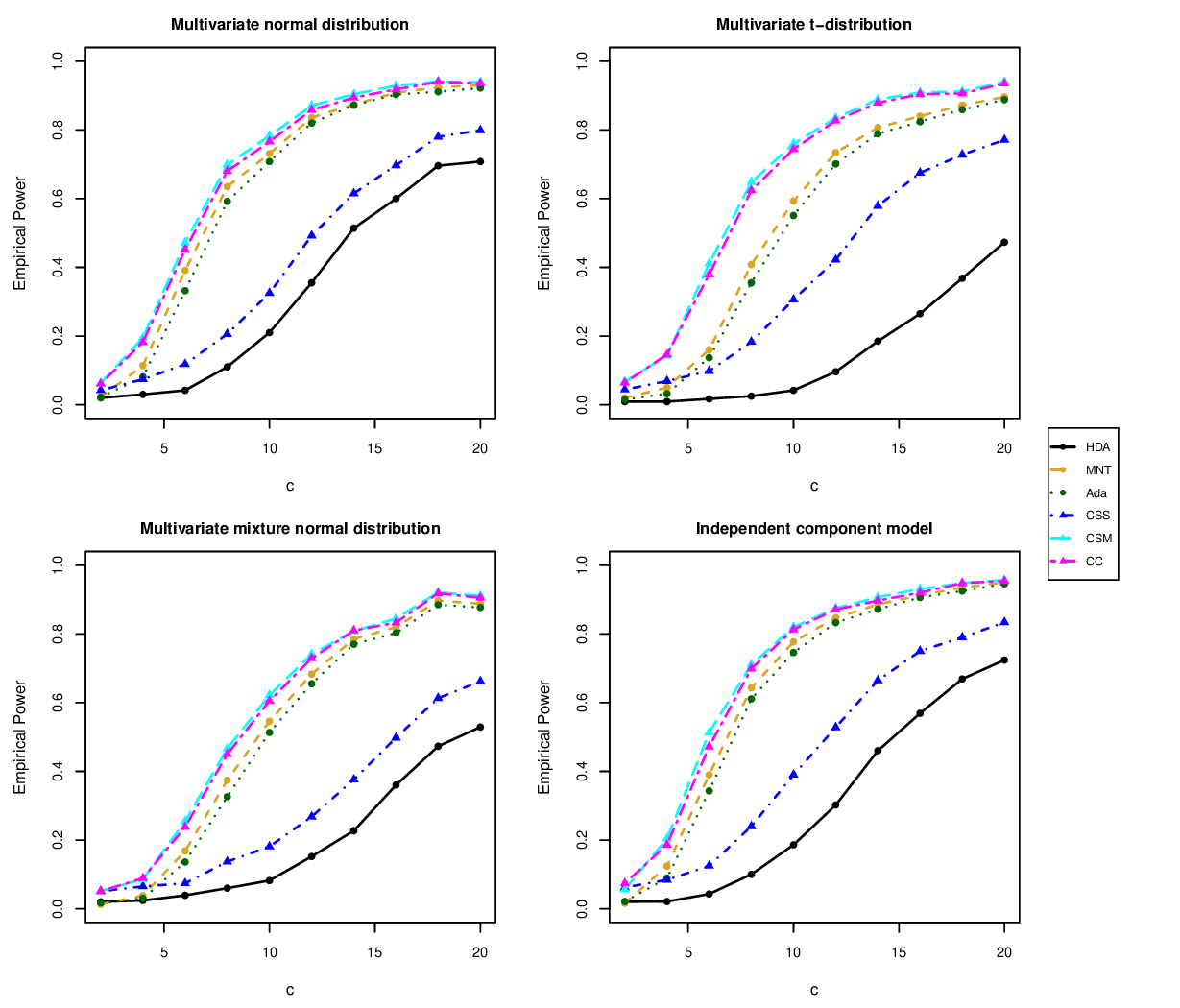}}
		\caption{Power performance against signal strengths $c$ for Example \ref{ex1}. \label{powersex1}}
\end{figure}

\begin{figure}[!ht]
	\centering
	{\includegraphics[width=1\textwidth]{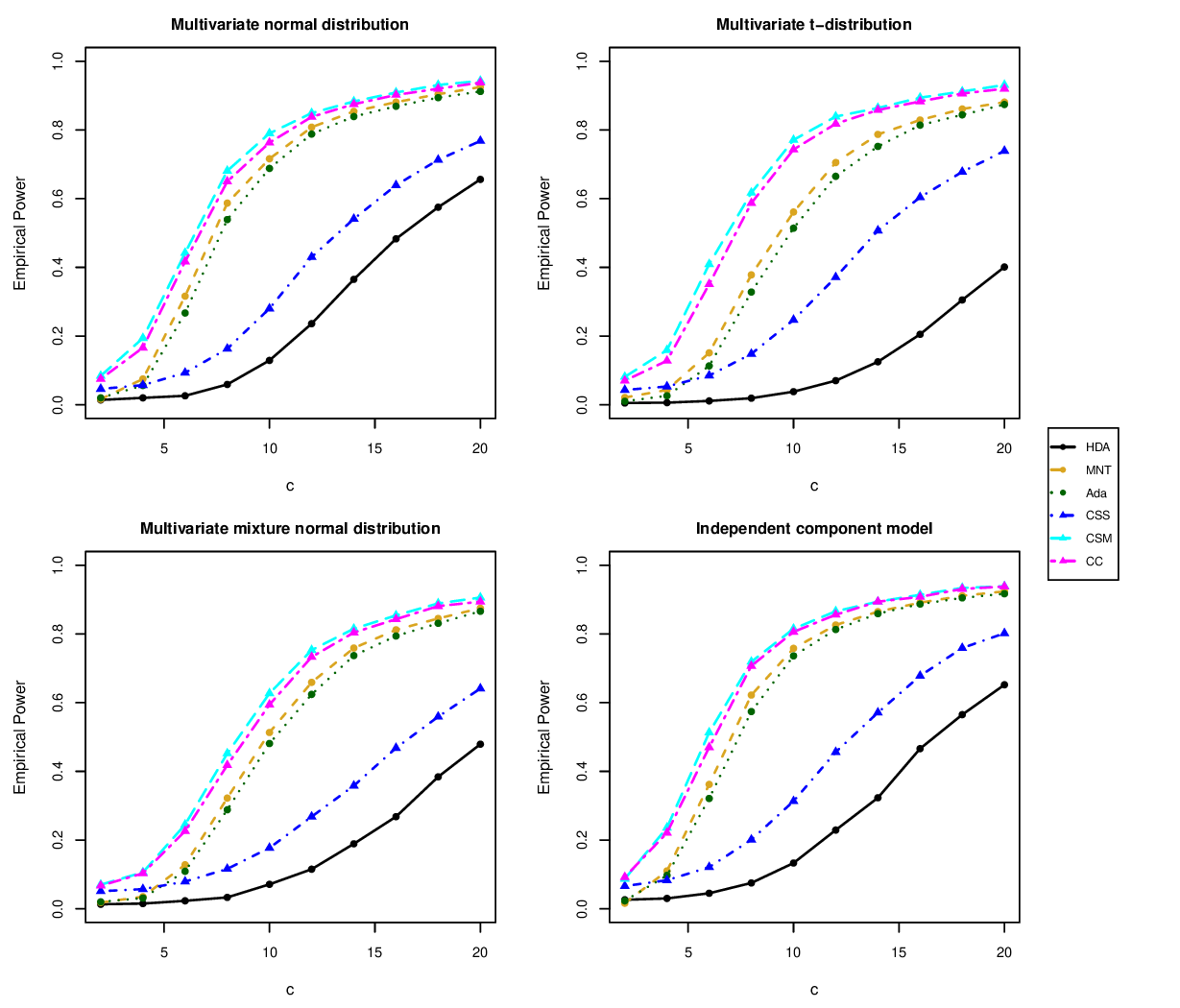}}
		\caption{Power performance against signal strengths $c$ for Example \ref{ex2}. \label{powersex2}}
\end{figure}

\begin{figure}[!ht]
	\centering
	{\includegraphics[width=1\textwidth]{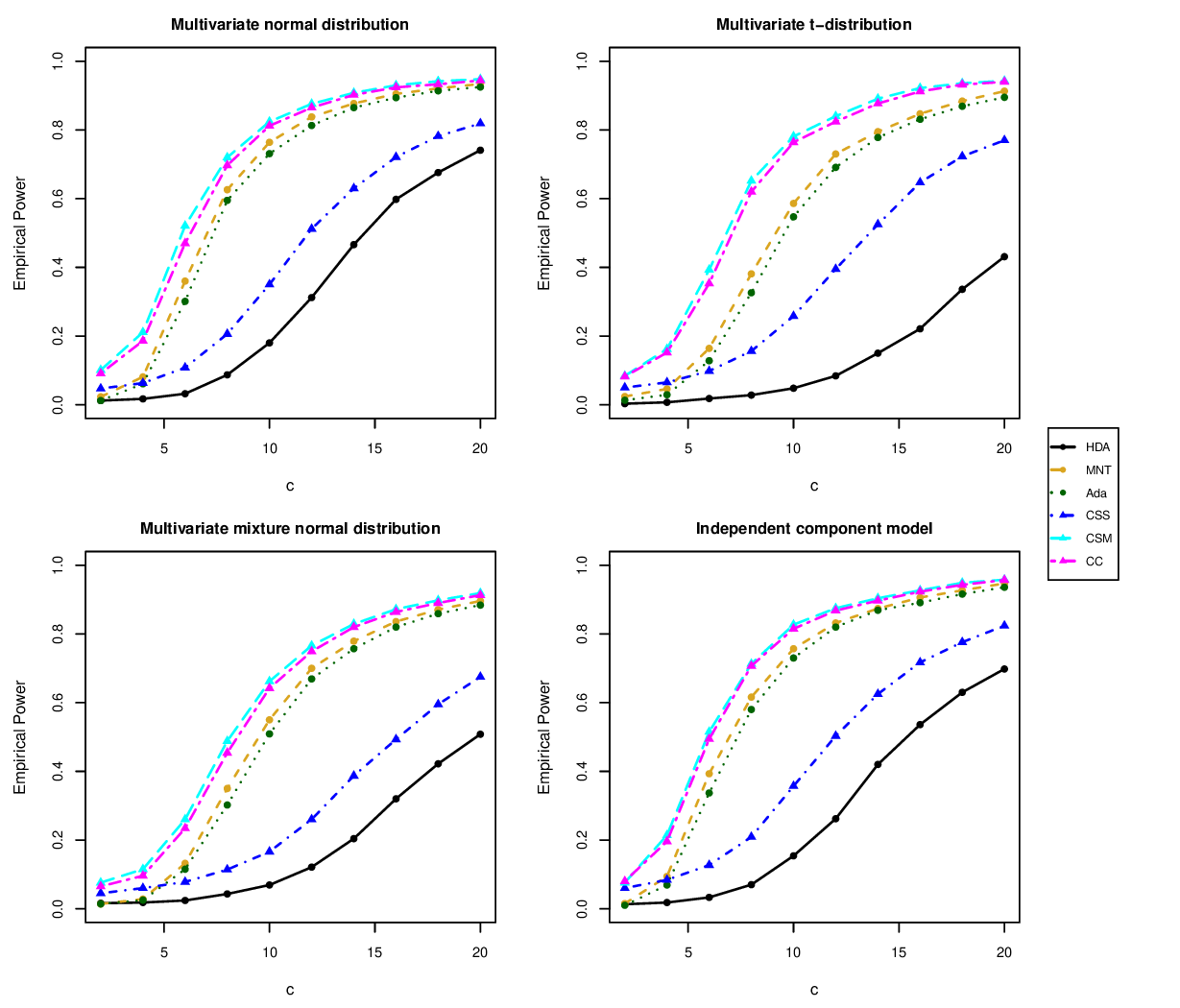}}
		\caption{Power performance against signal strengths $c$ for Example \ref{ex3}. \label{powersex3}}
\end{figure}

To investigate the empirical power of the proposed CSM and CC tests, we consider the alternative hypothesis where a subset of assets $\mathcal{S} \subset \{1, \dots, N\}$ exhibits non-zero intercepts. Let $|\mathcal{S}| = s$, where the indices in $\mathcal{S}$ are drawn uniformly at random from $\{1, \dots, N\}$. Following the signal strength specification in the existing literature \citep[e.g.,][]{ma2020testing}, we generate $\alpha_i$ independently from a uniform distribution $\mathcal{U}(0, c\sqrt{\log N / (Ts)})$ for each $i \in \mathcal{S}$, and set $\alpha_{it} = \alpha_i/T$. For $i \notin \mathcal{S}$, the intercepts are maintained at $\alpha_{it} = 0$. In all simulations, the dimensions are fixed at $(N, T) = (400, 350)$. We evaluate the sensitivity of the tests to both the sparsity level and the signal strength as follows:
\begin{itemize}
	
	\item[\textbf{(1)}] \textbf{Varying Sparsity:} To examine the impact of signal sparsity on test performance, we fix the signal strength at $c = 20$ and vary $s \in \{8, 16, \dots, 80\}$. Figures \ref{powerex1}-\ref{powerex3} summarize the empirical power across four error distributions for Examples 1-3, respectively.
	\item[\textbf{(2)}] \textbf{Varying Signal Strength:} We consider a highly sparse setting with $s = 2$ and vary the signal strength $c \in \{2, 4, \dots, 20\}$. The corresponding power curves are presented in Figures \ref{powersex1}-\ref{powersex3}.
\end{itemize}

Several key findings emerge from the simulation results. First, the proposed CC test consistently exhibits superior power performance across the entire range of $s$ and $c$, effectively dominating the competing procedures in nearly all configurations. Second, in sparse signal settings, the proposed CSM test outperforms the MNT test (the benchmark max-type test), while the CC test maintains a clear advantage over the Ada test (the benchmark adaptive test). These results underscore the robustness and adaptive nature of our proposed testing framework, particularly when dealing with heavy-tailed error distributions and varying signal structures.

   The simulation results indicate that the proposed CC test exhibits robust performance within the context of conditional time-varying factor pricing models, frequently outperforming the Ada test in the scenarios considered. By utilizing a spatial-sign-based construction, the CC test achieves improved power compared to existing methodologies. While conventional tests may experience power loss when error distributions exhibit thick tails, our procedure maintains higher efficiency and sensitivity to $\alpha$ deviations. Furthermore, the CC test demonstrates strong adaptivity across a diverse range of signal configurations, effectively capturing mispricing signals in both sparse and dense settings. These findings suggest that the CC test provides a reliable and versatile framework for conducting asset pricing tests in high-dimensional, time-varying environments.

\section{Real data application}\label{Emp}
To assess the performance of the proposed tests in a high-dimensional setting, we examine the constituent returns of the Standard \& Poor's 500 index, utilizing the same dataset
 as in \cite{ma2020testing}. The sample comprises weekly returns spanning from January 8, 2010, to August 25, 2017 ($T = 399$). To maintain a balanced panel and account for time-varying index composition, we restrict our focus to $N = 442$ securities that remained in the index throughout the entire sample period. We characterize the risk-return profile of these assets using the Fama-French three-factor model, formulated as:
\begin{align}\label{3f}
	Y_{it} = r_{it} - r_{ft} = \alpha_i + \beta_{i1} (r_{mt} - r_{ft}) + \beta_{i2} SMB_t + \beta_{i3} HML_t + \epsilon_{it},
\end{align}
where $Y_{it}$ denotes the excess return of security $i$ at time $t$, and $r_{mt} - r_{ft}$ represents the market risk premium. Time series data for the risk-free rate and the Fama-French factors were sourced from Kenneth French‘s data library. The one-month U.S. Treasury bill rate is used as the risk-free rate ($r_{ft}$), while the market return ($r_{mt}$) is proxied by the value-weighted return of all NYSE, AMEX, and NASDAQ stocks from CRSP. The size ($\text{SMB}_t$) and value ($\text{HML}_t$) factors are constructed following the standard methodology: $\text{SMB}_t$ represents the average return difference between three small and three big portfolios, while $\text{HML}_t$ is the average return difference between two value and two growth portfolios, all using stocks listed on the NYSE, AMEX, and NASDAQ. Finally, $r_{it}$ denotes the return of security $i$ at time $t$.

The primary objective is to test the validity of the factor pricing model by examining whether the pricing errors (alphas) are jointly zero. Specifically, we test the hypothesis: 
\begin{align*}
	H_0: \alpha_1 = \dots = \alpha_N = 0 \quad \text{versus} \quad H_1: \exists~i \in \{1, \ldots, N\} \text{ s.t. } \alpha_i \neq 0.
\end{align*}
For the full sample of length 399, the $p$-values for the HDA, MNT, Ada, CSS, CSM, and CC tests are 0, $2.42 \times 10^{-5}$, $1.71 \times 10^{-8}$, 0, $4.77 \times 10^{-9}$, and $1.04 \times 10^{-8}$, respectively. These results yield strong evidence to reject the null hypothesis, suggesting that the three-factor model does not fully account for the cross-sectional variation in returns over the entire sample period.

\begin{figure}[htbp]
	\centering
	\subfloat[$h=276$]{\includegraphics[width=0.85\textwidth]{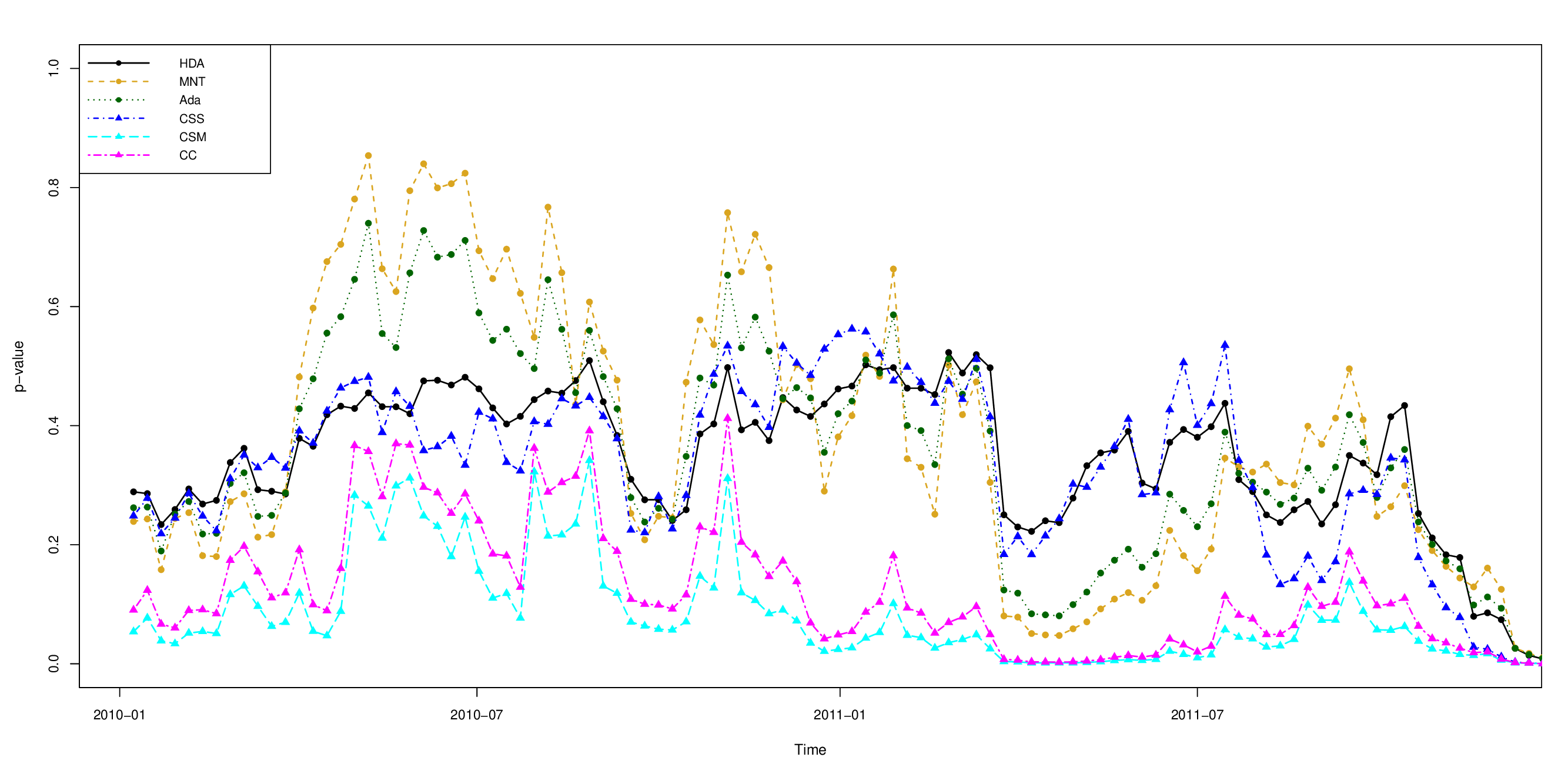}}\\
	\subfloat[$h=288$]{\includegraphics[width=0.85\textwidth]{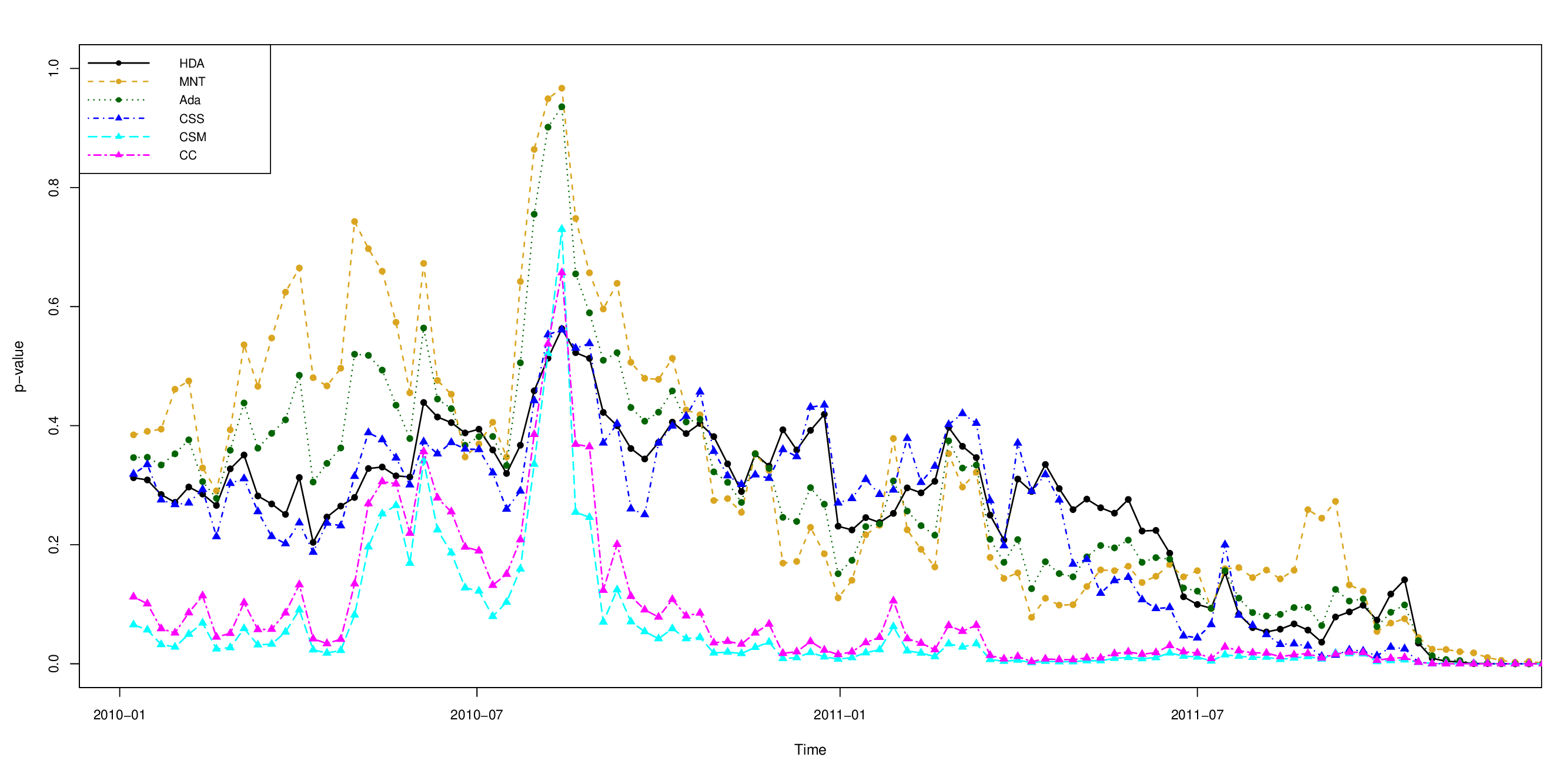}}\\
	\subfloat[$h=300$]{\includegraphics[width=0.85\textwidth]{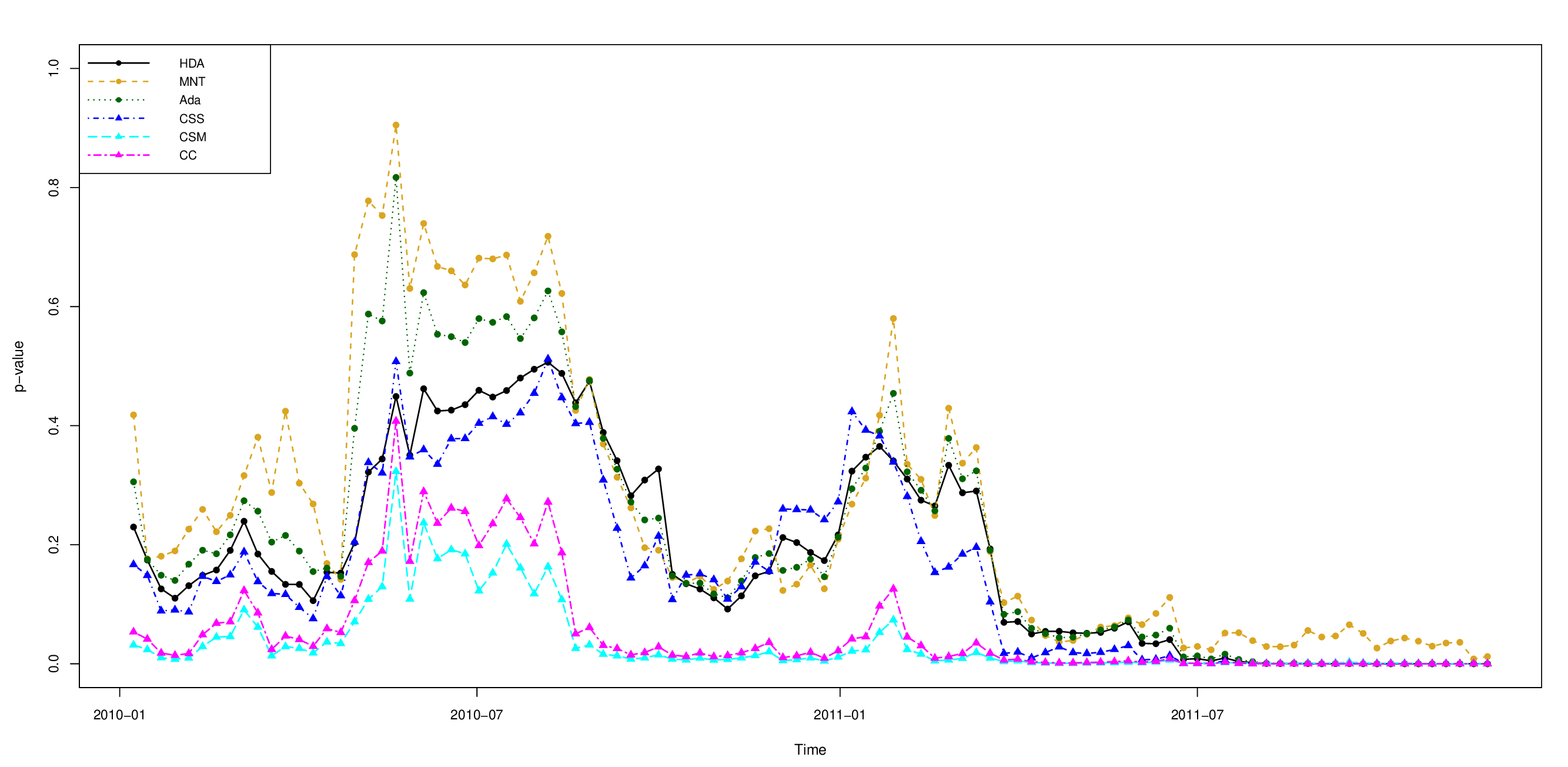}}
		\caption{$p$-value sequence of each test from 2010 to 2017 for US data. \label{figdata}}
\end{figure}

To account for potential structural instability or time-varying parameters, we implement a rolling window analysis. Specifically, the testing procedures are applied to each of the $399-h+1$ overlapping windows of length $h$, for $h \in \{276, 288, 300\}$. Table \ref{t3} reports the rejection ratios at the $\gamma = 0.01$ and $0.05$ significance levels, while Figure \ref{figdata} illustrates the sequential evolution of $p$-values for each test. Consistent with our theoretical and simulation evidence, several empirical findings are noteworthy. First, as summarized in Table \ref{t3}, spatial-sign-based tests consistently outperform their least-squares counterparts, demonstrating superior power and robustness in the presence of heavy-tailed financial data. Second, the proposed CC test achieves the most favorable performance across the majority of window specifications, underscoring its reliability in high-dimensional empirical settings.

\begin{table}[!ht]
	\begin{center}
		\caption{\label{t3}  The rejection ratios of the six tests with three different window lengths.}
		\vspace{0.5cm}
		\renewcommand{\arraystretch}{0.8}
		\setlength{\tabcolsep}{5pt}{
			\begin{tabular}{ccccccccc}
				\hline \hline
					$h$& HDA&MNT&Ada&CSS&CSM&CC\\ \hline
				& \multicolumn{6}{c}{$\gamma=0.01$}             \\
				276 & 0.171 & 0.130 & 0.171 & 0.187 & 0.309 & 0.236 \\
				288 & 0.153 & 0.108 & 0.144 & 0.180 & 0.360 & 0.324 \\
				300 & 0.212 & 0.010 & 0.192 & 0.273 & 0.394 & 0.364 \\\hline
				& \multicolumn{6}{c}{$\gamma=0.05$}             \\
				276 & 0.187 & 0.187 & 0.187 & 0.203 & 0.496 & 0.431 \\
				288 & 0.162 & 0.153 & 0.153 & 0.297 & 0.595 & 0.514 \\
				300 & 0.333 & 0.202 & 0.323 & 0.364 & 0.737 & 0.667\\ \hline
		\hline
		\end{tabular}}
	\end{center}
\end{table}

\section{Conclusion}\label{conclusion}

This paper develops a robust inferential framework for $\alpha$ testing within the context of high-dimensional, time-varying factor pricing models. We introduce a spatial-sign-based max-type test specifically designed to detect sparse mispricing signals. By establishing the Bahadur representation and Gaussian approximation for spatial median estimators, we derive the asymptotic null distribution and establish the consistency of the proposed test statistic.A primary theoretical result of this study is the proof of asymptotic independence between our spatial-sign max-type statistic and existing sum-type benchmarks. This independence facilitates the construction of a robust Cauchy combination (CC) test that provides a unified and adaptive approach for both sparse and dense alternative hypotheses. While existing robust methods in the literature provide a foundation for inference in these settings, the CC test significantly improves detection power by leveraging the geometric information of spatial signs and the flexibility of the Cauchy combination rule.Numerical results and empirical evidence underscore the advantages of the CC test, showing marked improvements in statistical efficiency under heavy-tailed distributions, a pervasive feature in high-dimensional equity markets.

\bibliographystyle{apa}\bibliography{Refers}

\end{document}